\documentclass[11pt]{article}
\usepackage{amsmath, amssymb, amsthm, float, fullpage, graphicx, multirow,  multicol, parskip,   setspace, subcaption}
\usepackage{tikz}

\usepackage{hyperref}
\usepackage{natbib}

\definecolor{PennRed}{RGB}{152, 30 50}
\definecolor{PennBlue}{RGB}{0, 44, 119}
\definecolor{PennGreen}{RGB}{94, 179,70}
\definecolor{PennViolet}{RGB}{141, 76, 145}
\definecolor{PennSkyBlue}{RGB}{14, 118, 188}
\definecolor{PennOrange}{RGB}{243, 117, 58}
\definecolor{PennBrightRed}{RGB}{223,82, 78}

\hypersetup{pdfborder = {0 0 0.5 [3 3]}, colorlinks = true, linkcolor = PennBrightRed, citecolor = PennSkyBlue}

\DeclareMathOperator*{\argmin}{arg\,min}

\title{Protocol for an Observational Study on the Effects of Playing Football in Adolescence on Mental Health in Early Adulthood}
\author{Sameer K. Deshpande\thanks{CSAIL, MIT. Corresponding author. Email: \nolinkurl{sameerd@alum.mit.edu}}, Raiden B. Hasegawa\thanks{The Wharton School, University of Pennsylvania}, Jordan Weiss\thanks{Population Studies Center, University of Pennsylvania}, Dylan S. Small\footnotemark[2]}
\date{9 November 2018}

\begin{document}
\maketitle

\def\C{\mathbb{C}}
\def\R{\mathbb{R}}
\def\Q{\mathbb{Q}}
\def\Z{\mathbb{Z}}
\def\N{\mathbb{N}}

\def\P{\mathbb{P}}
\def\E{\mathbb{E}}

\def\bR{\mathbf{R}}
\def\bZ{\mathbf{Z}}
\def\bX{\mathbf{X}}
\def\br{\mathbf{r}}
\def\bx{\mathbf{x}}

\begin{abstract}
\singlespacing{
More than 1 million students play high school American football annually, but many health professionals have recently questioned its safety or called for its ban. 
These concerns have been partially driven by reports of chronic traumatic encephalopathy (CTE), increased risks of neurodegenerative disease, and associations between concussion history and later-life cognitive impairment and depression among retired professional football players. 

A recent observational study of a cohort of men who graduated from a Wisconsin high school in 1957 found no statistically significant harmful effects of playing high school football on a range of cognitive, psychological, and socio-economic outcomes measured at ages 35, 54, 65, and 72. 
Unfortunately, these findings may not generalize to younger populations, thanks to changes and improvements in football helmet technology and training techniques. 
In particular, these changes may have led to increased perceptions of safety but ultimately more dangerous styles of play, characterized by the frequent sub-concussive impacts thought to be associated with later-life neurological decline. 

In this work, we replicate the methodology of that earlier matched observational study using data from the National Longitudinal Study of Adolescent to Adult Health (Add Health).
These include adolescent and family co-morbidities, academic experience, self-reported levels of general health and physical activity, and the score on the Add Health Picture Vocabulary Test. 
Our primary outcome is the CES-D score measured in 2008 when subjects were aged 24 -- 34 and settling into early adulthood.
We also examine several secondary outcomes related to physical and psychological health, including suicidality. 
Our results can provide insight into the natural history of potential football-related decline and dysfunction.

\vspace{0.1in}
}
\end{abstract}

\newpage
\section{Background and Motivation}
\label{sec:background_motivation}

American football is the largest participation sport in U.S. high schools -- more than 1 million boys played high school football in 2014. 
Playing football involves many hits to the head and is associated with a high rate of concussion of the brain. 
Football is a leading cause of concussion among adolescents and young adults \citep{Guskiewicz2000, Gessel2007}, accounting for nearly half of all sports-related concussions \citep{Marar2012}. 
These estimates may, in fact, underestimate the true prevalence of football-related head injury, as evidence suggests that as many of a half of all injuries may go unreported \citep{McCrea2004}.
Professional football players are 3 times more likely to die of a neurodegenerative disease than the general U.S. population \citep{Lehman2012}.
In a recent study of former football players who have donated their brains for posthumous study, \citet{Mez2017} reported that 110 of 111 former professional players had chronic traumatic encephalopathy (CTE), a neurodegenerative disease resulting from repetitive brain trauma whose symptoms may include memory loss, aggression, confusion, and depression \citep{Breslow2015}.
While most cases of CTE and its symptoms have been documents in older individuals, there is growing evidence that CTE can affect individuals in emerging and early adulthood.
For instance, \citet{Mez2017} also reports CTE in 48 out of 53 former collegiate players as well as in 3 out of 14 former high school players. 
Even more worryingly, CTE or early signs of it has been identified in the brains of several collegiate players who committed suicide in their early-20's while participating in college football \citep[see, e.g.][]{Schwarz2010, Bever2014, Bishop2018}.
Concerns over the long- and short-term cumulative effects of football-related head trauma have led several physicians to question the safety of youth football\citep{Bachynski2016, Pfister2016}, with some calling for its outright ban \citep{MilesPrasad2016}.

While the rates of CTE reported in \citet{Mez2017} and earlier studies are rather concerning, nearly all of the studies relied on convenience samples of donated brains.
As a result, these studies are all potentially affected by referral bias.
One exception is \citet{Bieniek2015}, which examined a larger, non-selected brain bank and found higher rates of CTE among former athletes than nonetheless. 
Though it is not as affected by referral bias, this study may be affected by information bias \citep{Rothman2008}, as it determined athletic participation by querying only medical records, potentially underestimating the true rate of participation.
Without population-level samples, it is difficult to estimate the base rate of dysfunction among professional players, much less among youth players. 
It is harder still to determine the causal effect of playing football on outcomes later in life. 

In theory, the ``gold standard'' approach for establishing such an effect to randomly divide students to two groups, one assigned to play football and one assigned to not play football, and to track the health outcomes of each group over several years.
Such a study, of course, is highly impractical and unethical and we instead must rely on observational data.
The main challenge in estimating the effect of a exposure (in this case, playing football) is that the exposed and control groups are generally not comparable prior to treatment. 
As a result, any observed difference in outcomes between the two groups is not immediately attributable to the exposure itself.
Arguably worse would be a harmful effect of football to remain undetected due to lack of control of important confounders. 
For instance, a population-level study \citet{Savica2012} found that high school football players were not at a higher risk of neurodegenerative disease than their male classmates who did not play football.
Unfortunately, they were unable to control for important confounders like family background, general health, or adolescent IQ \citep{Der2009, Rogers2009, Everson-Rose2003, Kaplan2001}.
While it certainly may be the case that playing football truly had no harmful effect, it is unclear whether differences in confounders ``washed away'' this effect.
A common statistical solution to this problem is to first match each treated subject to control subjects along several baseline covariates and then to compare the outcomes within each matched set.

In a recent paper, \citet{Deshpande2017} used data from the Wisconsin Longitudinal Study (WLS) to study the long-term cognitive and mental health consequences of playing high school football among a cohort of men who graduated from a Wisconsin high school in 1957. 
After matching football players to control subjects along a range of important baseline covariates, they found no statistically significant harmful effects of playing high school football on a several cognitive test scores and CES-D scores at ages 54, 65, and 72. 
While interesting, their findings may not generalize to younger cohorts of high school football players, due to changes in safety equipment and style of play.

In this work, we take a similar analytic approach to \citet{Deshpande2017} using data from the National Longitudinal Study of Adolescent to Adult Health (Add Health), a nationally representative longitudinal sample of United States middle and high school students.
Motivated in large part by reports of CTE and suicide among individuals in their early 20's \citep{Schwarz2010, Bever2014, Bishop2018}, we mainly focus on the effect of playing middle- or high-school football on mental health in early adulthood.
We specifically study the effect of playing school football on the score on the Center for Epidemiological Studies Depression Scale (CES-D) measured in Wave IV of the study in 2008 when subjects were between the ages of 24 and 32, after adjusting for a range of baseline variables measured in Wave I during the 1994-95 school year. 
We also study the effect of playing football on several secondary outcomes related to substance abuse, general health, and personality. 

Broadly, we conduct a matched observational study, in which we partition the football players and controls into smaller subgroups which are relatively homogeneous along a range of baseline covariates.
We then compare the outcomes within each matched set, after adjusting for residual imbalances in the distribution of these baseline covariates between the football players and controls.
To guard against concern that controls who play non-collision sports like basketball or swimming and controls who do not play any sports may systematically differ in unmeasured ways, we consider four comparisons:  football players compared to all controls, football players compared to sport-playing controls, football players compared to non-sport controls, and sport-playing controls compared to non-sport playing controls.
For each comparison, we partition the relevant subjects into smaller subgroups which are relatively homogeneous along a range of baseline covariates using propensity score matching.
We are then able to compare outcomes within these matched sets in order to perform inference about the treatment effect, after appropriate adjustment for residual imbalances in these covariates. 

The rest of this protocol is organized as follows.
In Section~\ref{sec:add_health}, we briefly describe the Add Health study and outline the eligibility and exclusion criteria and outcomes we considered in the proposed study. 
We review propensity score matching and randomization inference with covariate adjustment in Sections~\ref{sec:methods}.
Next, we describe the resulting matched sets for each comparison in Section~\ref{sec:results} and conclude with a discussion of the strengths and limitations of the proposed study in Section~\ref{sec:discussion}.

\section{The Add Health Dataset}
\label{sec:add_health}

The National Longitudinal Study of Adolescent to Adult Health (Add Health) is a nationally representative and longitudinal study of American adolescents. 
Study participants were drawn from a probability sample of 80 high schools and 52 middle schools which were representative of the US (grades 7-12) in 1994-1995 \citep{Harris2013}
More than 90,000 participants were invited to complete an in-school survey during the baseline years 1994-1995, from which a random sample of over 20,000 students were selected to complete an in-home survey in 1995 (Wave I). 
Participants were followed-up in 1996 (Wave II), 2001 -- 02 (Wave III), and 2008 (Wave IV) to provide data on sociodemographic characteristics, household characteristics, school participation, and extracurricular activities, including participation in contact sports. 
For more detailed information on the Add Health study and its design, please see \citet{Resnick1997} and \citet{Harris2013}.

Our single primary outcome is the Center for Epidemiological Studies Depression Scale (CES-D) score \citep{Radloff1977} measured in 2008 during the Wave IV survey when subjects were between 24 and 32 years old.
We will also examine the effect of playing football on several secondary outcomes, all of which were measured during Wave IV and are listed below along with the corresponding Add Health variable names in parentheses.

\textbf{General Health Outcomes}
\begin{itemize}

\item{Binary indicator of being a daily smoker at Wave IV (C4VAR035)}
\item{Binary indicator of being physically active at Wave IV (C4VAR036)}
\item{Binary indicator of receiving a diagnosis of high blood cholesterol, triglycerides, or lipids after age 18 (H4ID5B, H4ID6B)}
\item{Binary indicator of receiving a diagnosis of high blood pressure or hypertension after age 18 (H4ID5C, H4ID6C)}
\item{Binary indicator of receiving a diagnosis of high blood sugar or diabetes after age 18 (H4ID5D, H4ID6D)}
\item{Binary indicator of receiving a diagnosis of heart disease after age 18 (H4ID5E, H4ID6E)}
\item{Binary indicator of receiving a diagnosis of migraine headaches after age 18 (H4ID5G H4ID6G)}
\item{Binary indicator of receiving a diagnosis of depression after age 18 (H4ID5H, H4ID6H)}
\item{Binary indicator of receiving a diagnosis of PTSD after age 18 (H4ID5I, H4ID6I)}
\item{Binary indicator of receiving a diagnosis of anxiety or panic disorder after age 18 (H4ID5J, H4ID6J)}
\item{Binary indicator of having seriously considered suicide in the past year (H4SE1)}
\end{itemize}

\textbf{Substance Dependence/Abuse Outcomes}
\begin{itemize}
\item{Binary indicator of nicotine dependence based on Fagerstrom scale (C4VAR017)}
\item{Binary indicator of a DSM4 lifetime diagnosis of alcohol abuse or dependence (C4VAR023)}
\item{Binary indicator of a DSM4 lifetime diagnosis of cannabis abuse or dependence (C4VAR028)}
\end{itemize}

\textbf{Personality Outcomes:}
\begin{itemize}
\item{Cohen Perceived Stress Scale (C4VAR001)} 
\item{Extraversion Personality Scale (C4VAR004)}
\item{Neuroticism Personality Scale (C4VAR005)}
\item{Agreeableness Personality Scale (C4VAR006)}
\item{Conscientiousness Personality Scale (C4VAR007)}
\item{Open to Experience/Intellect/Imagination Personality Scale (C4VAR008)}
\item{Anxious Personality Scale (C4VAR009)}
\item{Optimistic Personality Scale (C4VAR010)}
\item{Angry Hostility Personality Scale (C4VAR011)}
\end{itemize}

A key strength of the Add Health study is the rich collection of baseline covariates measured prior to (or concurrent with) exposure in the Wave I home interview.
This enables us to control for a very large number of potential confounders.
In addition to basic demographic variables, we identified several groups of baseline covariates that may covary with treatment status and also impact later-life health.
These include measures of general health, propensity for ``risky'' or ``delinquent'' behavior, patterns of adolescent substance abuse, exposure to violence, suicidality, exposure to ``protective factors'', view of the future (i.e. life plan), and measures of subjects' feelings and personality. 
A full list of the covariates considered is available in Appendix~\ref{app:baseline_covariates}.
We will divide our study subjects into several matched sets consisting of one treated subject and a variable number of controls in a way that balances the distribution of these covariates between treated and control groups.

\subsection{Eligibility and Exclusion Criteria}
\label{sec:eligibigility_exclusion_criteria}

In all, there are 5,780 men in the core Add Health sample, of whom 3,989 had completed the in-school questionnaire.
This questionnaire collected measurements on a range of topics including demographics, family background, educational and economic aspirations, and performance and behavior in school.
Additionally, it asked about participation in a range of school activities, including football and several other school sports.
Our initial pool of ``treated'' subjects were those men who indicated that they participated in or intended to participate in school football. 
Before proceeding, it is critical to stress that the Add Health dataset does not contain a direct measure of actual football participation or exposure to football-related head trauma.
This means that our treated group may in fact contain some non-compliers who, despite indicating that they intended to participate in football, ultimately did not. 
This may limit the applicability and generalizability of our eventual results.

Since our main interest is in the effects of playing football, we removed the 993 men in the core sample who participated in or intended to participate in at least one of field hockey, ice hockey, soccer and wrestling, as these sports are all associated with a high risk of repetitive head trauma. 
Included in these 993 excluded men were those who participated in ``Other sports.''

As part of the Wave I in-home interview conducted in 1995, subjects were asked several questions related to physical or functional limitations such as ``Do you have difficulty using your hands, arms, legs, or feet because of a permanent physical condition?'' and ``Do you have difficulty walking up 10 steps without resting?''
Based on the responses to these questions, we determined that an additional 119 men had some type of physical or functional limitation or disability.
These men were excluded from our study. 

Finally, 680 (140 football players, 540 controls) of the remaining 2,877 men were missing measurements of the primary outcome, the CES-D score measured in the 2008 in-home survey (Wave IV). 
On further inspection, we found that these subjects were missing a large majority of the secondary outcomes listed above that were also recorded in Wave IV. 
This would suggest that these individuals did not participate in the Wave IV survey. 
If football playing status affected the availability of 
If football playing status affects the availability of these results (e.g. if playing football increased the likelihood of dying younger or early onset of debilitating impairment so that subjects were unable to participate in the Wave IV study), any comparison of the treated and control group based on these outcomes will be biased.
To examine this possibility, we fit separate logistic regression models to predict the availability of each outcome listed above using the baseline covariates listed in Appendix~\ref{app:baseline_covariates} and treatment status. 
We found that playing football was not a significant predictor of whether a subject was missing any of the outcomes. 

Following these exclusions, we are left with 2,197 subjects, of whom 521 indicated that they participated in or intended to participate in football.
Of the 1,676 control subjects, 610 played a non-collision sport like basketball, swimming, or tennis, and the remaining 1,066 did not participate in any school sport.
Before proceeding, we stress that our data does not record whether or not subjects actually played school football. 
This may limit the applicability and generalizability of any of our results. 

\section{Methods}
\label{sec:methods}

The problem of estimating causal effects is, at its core, a problem of comparing potential outcomes \citep{Rubin1974}: to understand the effect that a particular treatment has on a particular individual, we must compare the outcome we would observed had he been given treatment to the outcome we would observe had he not been given the treatment.
Unfortunately, as \citet{Holland1986} notes, ``the fundamental problem of causal inference'' is that we only ever observe one of these potential outcomes, since at any point in time each individual receives either the treatment or control.
As a result, we must draw inference about causal effects by comparing the outcomes of treated and control individuals who may differ along many baseline measured or unmeasured covariates.
Intuitively, the more similar the distributions of these covariates are between the two groups, the more inclined we are to believe that the comparison of their outcomes uncovers the causal quantities of interest. 

It is well-known that randomizing the treatment assignment tends to \textit{balance} the distribution of covariates, both measured and unmeasured, between the control and treatment group.
In such settings, a direct comparison, say of the average outcomes observed for the treated and control groups, is a perfectly reasonable causal estimate.
In observational studies, however, when the treatment is not randomly assigned, there tends to be substantial imbalance between treated and control groups and such a comparison is decidedly less reasonable. 
Matching is a popular and widely applicable tool to overcome this hurdle by partitioning the study population into several \textit{matched sets} consisting of at least one treated and at least one control subject in such a way that the matched sets are aligned along all of the observed covariates. 
For a comprehensive and accessible review on the role of matching in causal inference, please see \citet{Stuart2010}. 

In this work, we construct matched sets containing one treated subject and a variable number of controls so that the distribution of the covariates listed in Appendix~\ref{app:baseline_covariates} are as similar as possible within matched sets. 
To set our notation, there are $I$ matched sets and within set $i,$ there are $n_{i} - 1$ control subjects and one treated subject.
We will let $Z_{ij} = 1$ if the $j^{\text{th}}$ subject in set $i$ was exposed and $Z_{ij} = 0$ otherwise, so that $\sum_{i = 1}^{n_{i}}{Z_{ij}} = 1.$
We will let $\bx_{ij}$ be the vector of observed baseline covariates for the $j^{\text{th}}$ individual in set $i.$
Each subject has two potential outcomes: the response $r^{(T)}_{ij}$ that would be observed if subject $j$ in matched set $i$ received treatment and $r^{(C)}_{ij}$ if he were assigned to control.
We only observe one of $r^{(C)}_{ij}$ and $r^{(T)}_{ij},$ precluding direct calculation of the effect caused by giving treatment instead of control, $r^{(T)}_{ij} - r^{(C)}_{ij}.$
Let $R_{ij} = r^{(C)}_{ij}(1 - Z_{ij}) + r^{(T)}_{ij}Z_{ij}$ be the observed response for individual $j$ in matched set $i$.
We will denote the vectors of observed responses and treatment indicators $\bR$ and $\bZ,$ respectively.
We will also let $\br^{(C)}$ and $\br^{(T)}$ be the vectors of the potential outcomes so that $\bR = \br^{(C)}(\boldsymbol{1} - \bZ) + \br^{(T)}\bZ,$ where $\boldsymbol{1}$ is the vector of all 1's.
In this framework, $\bR$ and $\bZ$ are observed but $\br^{(C)}$ is not. 
Throughout, we assume an additive treatment model so that for all $i,j$, $r^{(T)}_{ij} = r^{(C)}_{ij} + \tau$ for a fixed constant $\tau.$

\subsection{Matching with a Variable Number of Controls}
\label{sec:matching}

An intuitive first attempt at matching would be to group treated subjects with those controls subjects whose baseline covariate are identical. 
When there are several baseline covariates, however, such a strategy is difficult if not impossible to implement.
Instead, we aim to create matched sets in such a way that for each covariate, the mean value of the covariate among the matched treated subjects is similar to the mean value of the covariate among the matched control subjects. 
To assess the suitability of a match, we look at the standardized difference in the mean of each covariate between treated and control groups. 

To construct balanced matched sets, we began by estimating the \textit{propensity score}, $e(\bx) = \P(Z = 1 | \bx)$, or the conditional probability of being assigned treatment ($Z = 1$) given the baseline covariates ($\bx$).
As \citet{Rosenbaum1983} show, the treatment assignment and covariates are conditionally independent given the propensity score so matching based on the propensity score tends to balance the 
distribution of the observed covariates between treated and control groups. 
Rather than matching strictly on the estimated propensity score (see Section~\ref{sec:propensity_score} for details), we instead match based on propensity score-calipered rank-based Mahalanobis distance between the observed covariates of each pair of treated and control subject.
The combination of a propensity caliper and Mahalanobis distance strives to achieve a good compromise between overall covariate balance and closeness of the covariates of matched subjects \citep[see, e.g.][]{GuRosenbaum1993}. 

We divided our subjects into three groups, based on what grade they were in when they completed the in-school questionnaire.
Specifically, we we grouped seventh and eight graders together, ninth and tenth graders together, and eleventh and twelfth graders together.
Such stratification ensures that we do not match, for instance, a seventh grade student with a twelfth-grader. 
Within each stratum, we performed variable ratio matching \citep{MingRosenbaum2000, Pimentel2015} using our calipered distance matrix. 

The variable ratio matching procedure employed within each grade-level stratum works as follows.
We take $K = 15$ to be the maximum number of controls that we will allow to be matched to a single treated subject.
Then we define sets $S_{1} = (1/3, 1], S_{15} = [0, 1/16]$ and $S_{k} = (1/(k+2), 1/(k+1)]$ for $k = 2, \ldots, 14.$
For each $k = 1, \ldots, 15,$ we select all subjects whose estimated propensity scores fall into the interval $S_{k}$ and perform $1:k$-matching among these selected subjects.
Whenever there are more than $k$ times as many control subjects as exposed, some control subjects are dropped from the analysis.
Additionally, when there are more treated subjects and than control subjects in the selected set, we build pair matches, optimally discarding the extra treated subjects who are most dissimilar to the controls within the stratum under consideration \citep{Rosenbaum2012}.
We note that our matching procedure does not use the full set of study subjects; instead it attempts to optimally discard subjects for whom there are comparable subjects with the opposite treatment assignment. 

\subsection{Randomization Inference and Covariance Adjustment}
\label{sec:covariance_adjustment}

Once we create the matched sets, we are ready to compare the primary outcomes of the treated and control subjects within each matched set.
Though matching can help eliminate some bias from this comparison, by balancing covariates on average, there may still be additional remaining bias due to residual covariate imbalances.
One way to further reduce both the bias and variance stemming from residual covariate imbalance is to combine matching with covariate adjustment via regression \citep{Rosenbaum2002a, Hansen2004}.
We follow the general procedure introduced in \citet{Rosenbaum2002a}, which is outlined below, for testing $H_{0}: \tau = \tau_{0}.$

Under this null, we know that for treated individuals their potential outcome under control is simply $R_{ij} - \tau_{0}.$
Notice further that under the null and our assumed additive treatment model, the vector of adjusted responses $\bR - \tau_{0}\bZ$ is fixed and does not vary with the treatment assignment. 
We then center the adjusted responses and covariates within matched sets, yielding aligned, adjusted responses
$$
\tilde{R}_{ij} = (R_{ij} - \tau_{0}Z_{ij}) - n_{i}^{-1}\sum_{k = 1}^{n_{i}}{(R_{ik} - \tau_{0}Z_{ik})}.
$$
and aligned covariates
$$
\tilde{\bx}_{ij} = \bx_{ij} - n_{i}^{-1}\sum_{k = 1}^{n_{i}}{\bx_{ik}}.
$$
Next, we regress $\tilde{\bR}$ onto $\tilde{\bX}$ and compute the vector of residuals $\boldsymbol{\varepsilon}.$
We finally perform a permutational t-test using these residuals. 

Since there is only ever at most one value of $\tau_{0}$ for which the null $H_{0}: \tau  = \tau_{0}$ is true, we are permitted to test many such hypotheses at a fixed significant level, say $\alpha = 0.05,$ using the above procedure.
By tracking those values of $\tau_{0}$ for which we do not reject the null, we may invert the test and form a $100\times(1-\alpha)$ confidence set for the treatment effect $\tau.$
This inversion, while straightforward in principle, is complicated by the fact that our testing procedure is not necessarily monotonic.
That is, if we reject the null at some $\tau_{0},$ then we will reject at every $\tau_{0}' > \tau_{0}.$
This is largely thanks to the non-linear nature of the BART fit. 
Nevertheless, to approximate a 95\% confidence region, we will test along a fine grid of pre-specified $\tau_{0}$ values.
In particular, for the continuous outcomes considered, we will let this grid contain weak, medium, and strong effect sizes, as defined by \citet{Cohen1988}. 

\subsection{Sensitivity Analysis}
\label{sec:sensitivity_analysis}

Controlling for observed baseline covariates through matching is designed to eliminate bias in treatment assignment by balancing the distribution of observable potential confounders between the treated and control groups. 
However, it cannot ensure balance of unobserved confounders unless they are highly correlated with observed confounders. 
Conditional on the match, we let $\Gamma$ bound the odds ratio of treatment for any pair in a matched set. 
For each marginally significant outcome in both the primary and secondary analysis we will report the $\Gamma$ at which the result is sensitive (i.e. the $\Gamma$ at which the result becomes insignificant). 
This is known as a sensitivity analysis \citep{Rosenbaum2002} and allows us to quantitatively assess how sensitive the results are to bias in the treatment assignment due to the observational nature of the study.  
Larger values of $\Gamma$ provide greater evidence for the study's causal conclusions. 
Sensitivity analyses for the continuous primary and secondary outcomes will be performed with the \texttt{sensitivitymv} package in \texttt{R}.  
The sensitivity analysis will be performed on the residuals after regressing the outcomes on the covariates -- this is the covariance adjustment procedure suggested by \citep{Rosenbaum2002a}. 
For the binary secondary outcomes, we will use a sensitivity analysis for testing the null hypothesis of no treatment effect using the Mantel-Haenszel test \citep[Section 4.2]{Rosenbaum2002}.

\subsection{Choice of Covariance Adjustment Procedure}
\label{sec:bart}

Before proceeding, we note that \citet{Rosenbaum2002a} does not prescribe the specific form of the regression used to produce $\boldsymbol{\varepsilon}$, noting that while the ``specific fitting algorithm used is of practical importance,'' the choice does not affect ``the logical structure of the argument.''
While using ordinary least squares is a straightforward option, we might reasonable suspect that there is a non-linear dependence between the adjusted responses and the covariates that involves complex higher-order interactions.
In light of this, we actually produce the residuals using the Bayesian Additive Regression Tree (BART) approach of \citet{Chipman2010} to carry out our regression adjustment. 
At a high-level, BART flexibly models the relationship between predictors and outcomes by training several shallow regression trees and then summing the predictions from each tree. 
Since its introduction, BART has demonstrated considerable success in automatically identifying non-linear relationships, interactions, and discontinuities, without requiring us to specify a parametric relationship between inputs and output \textit{a priori}.
Please see \citet{Chipman2010} for more details about the BART algorithm. 

Over the last several years, there has been growing interest in using BART for causal inference \citep[see, e.g.][]{Hill2011a, GreenKern2012, Hill2013, Sivaganesan2017, Hahn2017}. 
Notable among these are \citet{Hill2011a} and \citet{Hahn2017} who use BART to flexibly model the two potential response surfaces $\E[r^{(C)} | \bx]$ and $\E[r^{(T)} | \bx].$ 
This approach has shown great promise in estimating individual treatment effects in a number of independent, simulated studies \citep{Dorie2018}.
We pause here to highlight a few distinctions between our use of BART and these approaches.
First, we use BART simply as a means to obtain an \textit{algorithmic} fit and our inference about the constant treatment effect $\tau$ does not depend on the Gaussian error model underlying the standard implementation of BART. 
This is in marked contrast to \citet{Hill2011a} and \citet{Hahn2017} who rely on this stochastic model as the basis of their inference about causal effects. 
Secondly, sensitivity analyses for $m$-tests like the permutational t-test we use has been well-established \citep[see][]{Rosenbaum2007}.
To the best of our knowledge, there are no extent procedures for assessing the sensitivity of \citet{Hill2011a} and \citet{Hahn2017}'s results to unmeasured confounding.

\subsection{Estimating the Propensity Score}
\label{sec:propensity_score}

In order to construct the match described in Section~\ref{sec:matching}, we must estimate the propensity score $e(\bx) = \P(Z = 1 | \bx),$ or conditional probability of treatment as a function of the baseline covariates $\bx.$ 
Following a recommendation of \citet{Gelman2008}, we re-center and re-scale the continuous covariates to have mean zero and standard deviation 0.5.

\citet{DehejiaWahba1999} recommend starting with the simple linear-logistic model
\begin{equation}
\label{eq:naive_propensity}
\log{\left(\frac{e(\bx)}{1 - e(\bx)}\right)} = \bx^{\top}\beta
\end{equation}
and estimating the propensity scores via maximum likelihood.
From there, we can construct a match as described above and assess the covariate balance by computing the post-match standardized difference in means between control and treated group. 
Should this match not adequately balance the covariates, \citet{DehejiaWahba1999} recommend introducing additional interaction and non-linear terms to the equation above, re-estimating the propensity scores, and assessing the balance.
This process continues until adequate balance is achieved.
While this process is intuitive, it is not especially viable when the number of pre-treatment covariates $p$ is large.
In such settings, maximum likelihood estimation of $\beta$ in Equation~\eqref{eq:naive_propensity} is often not feasible, due to issues of perfect separation. 
Moreover, even when the MLE $\hat{\beta}_{MLE}$ is non-degenerate, the propensity score estimates tend to over-fit the data; please see \citet{Hill2011b} and \citet{Spertus2017} for a more in-depth discussion of the challenges of estimating high-dimensional propensity scores. 

Instead, we consider four different estimates of the propensity score and pick the one which yields the match that i) adequately balances the covariates between treated and control group and ii) discards the fewest subjects. 
We first consider the maximum likelihood estimate of the propensity score.
Next, rather than simply minimizing the negative log-likelihood corresponding to~\eqref{eq:naive_propensity}, we instead solve the penalized optimization problem:
$$
\hat{\beta}_{\lambda} = \argmin\left\{\sum_{i = 1}^{n}{-y_{i}\bx_{i}^{\top}\beta + \log{(1 + \exp(\bx_{i}^{\top}\beta)}} + \lambda \lVert \beta \rVert\right\}
$$
and plug $\hat{\beta}_{\lambda}$ instead of $\hat{\beta}_{MLE}$ into the formula for the propensity score. 
The $\ell_{1}$ penalty ensures that the estimate $\hat{\beta}_{\lambda}$ is sparse, so that only a handful of covariates are used to estimate the propensity score.
The sparsity of $\hat{\beta}_{\lambda}$ is controlled by the parameter $\lambda,$ which we select via cross-validiation to help protect against potential overfit. 

We next consider a standard Bayesian estimate of the propensity score formed by placing a standard multivariate normal prior $N_{p}(0, I)$ on $\beta$ and use a Markov Chain Monte Carlo (MCMC) simulation to generate draws from the posterior distribution of $\beta.$
We then plug in each of these posterior draws into the formula for the propensity score to generate draws from the induced posterior distribution of $e(\bx).$
We finally estimate $e(\bx)$ by its posterior mean. 
Under our scaling of the covariates, our prior reflects the belief that a two standard deviation increase in a single covariate, keeping all else fixed, is very unlikely to change the log-odds of treatment by an additive factor of 3. 
This informative prior regularizes our propensity score estimates and can help guard against overfit. 
We fit this model using Stan \citep{Stan}.

These three estimates of the propensity score are based on the model in Equation~\eqref{eq:naive_propensity}, which asserts that the log-odds of treatment depend on the covariates in a linear fashion. 
Prima facie, it is perhaps more realistic to believe that an individual's likelihood to play football is based on more complex interactions between the covariates.
This motivates our final propensity score estimate, which expresses the log-odds of treatment assignment as a sum of regression trees, which are fit using a variant of \citet{Chipman2010}'s BART procedure.
This procedure is rather attractive as it not only regularizes our the propensity score estimates but also does not require us to specify the functional form of $e(\bx).$
\citet{Spertus2017} has reported success in estimating causal effects using BART-based propensity score estimates.

\subsection{Multiple Comparison Groups}
\label{sec:multiple_comparison_groups}

Within the pool of control subjects, we find that 610 participated in a non-collision sport (e.g. basketball, tennis, track) while 1,066 participated in no school sports at all.
We might reasonably expect that the pool of sport-playing control subjects may differ substantially from non-participants along several dimensions related to personality, temperament, overall fitness, and lifestyle.
Such differences, especially along unmeasured dimensions, can introduce problematic confounding to the main comparison of football players and all control units.
At first glance, a natural approach to alleviate concern about such unmeasured confounding would be to exclude the controls who did not play sports at all and focus only on comparison football players with sport-playing controls.
Doing so, however, would effectively half our sample size and could result in a substantial decrease in power.
Besides a potential loss of power, removing the non-sport-playing controls limits the hypotheses we are able to probe. 
For instance, determining whether i) playing sports in general has a net positive effect on mental health in early adulthood but ii) football players fared worse, on average, than those who participated in non-collision sports requires comparing the football players to each subgroup. 

We follow \citet{Deshpande2017} and consider four comparisons in order: football players versus all controls, football players versus sport-playing control, football players versus non-sport-playing controls, and sport-playing controls non-sport-playing controls.
A convincing study would show consistent evidence (or lack thereof) of an exposure effect across the first three comparisons.
Moreover, comparability of the two control groups in the fourth comparison would suggest that the differences between the sport-playing and non-sport-playing controls did not significantly influence the outcome of interest \citep{Rosenbaum2002}.
Please see \citet{Rosenbaum1987} and \citet{Yoon2011} for a more detailed discussion about the use of multiple control groups. 

By using both control groups separately we systematically vary the unmeasured confounders of concern.
In order to preserve the increased power of using controls from both groups while still testing the treated group against each group separately we consider an ordered testing procedure which controls the family wise error rate (FWER) \citep{Rosenbaum2008}. 
In particular, we first test the null of no treatment effect using matched sets constructed with all controls.
If we reject this hypothesis at level $\alpha = 0.05$, we can conduct the same test separately to compare football players to each subgroup of controls.
If we reject both tests at level $\alpha = 0.05,$ we can then perform an equivalence test between the two subgroups of control.
Finally, if at any stage, we do not reject the hypothesis $H_{0}: \tau = \tau_{0}$, we stop the procedure.
This stopping rule guarantees FWER control at level $\alpha = 0.05.$
Please see \citet{Hasegawa2017} for a much detailed description and extensions of this procedure. 

To carry out this testing in order procedure, we build four separate matches, one for each comparison, and carry out the test of no treatment effect using the methods outlined above.
When comparing the sport-playing controls to the non-sport controls, we take the ``treatment'' to be playing a non-collision, non-football sport.

\subsection{Secondary Analyses}
\label{sec:secondary_analyses}

For continuous secondary outcomes, namely the scores on the personality scales listed in Section~\ref{sec:add_health}, we will use the same procedure as described in Section~\ref{sec:covariance_adjustment}.
Specifically, for each of these continuous outcomes, we will test the hypothesis of no treatment effect ($H_{0}: \tau = 0$) at the 95\% significance level.
For the binary secondary outcomes, we will run a conditional logistic regression and test the hypothesis of no treatment effect also at the 95\% significance level.
This will all be done within the testing-in-order procedure outlined in Section~\ref{sec:multiple_comparison_groups}.
That is, if we reject the null of no treatment effect when comparing football players to all controls at the 5\% level, then we will test the null of no treatment effect by comparing the football players to each subset of the controls. 

We will report both unadjusted p-values as well as p-values adjusted using the Benjamini-Hochberg procedure, if any of the unadjusted p-values falls below the 5\% threshold. 
Finally, we will report approximate 95\% confidence intervals for the estimated treatment effects, without adjusting for multiple testing.
In general, these confidence intervals will tend to be shorter as they only achieve marginal coverage rather than joint coverage. 

\section{Matching Results}
\label{sec:results}


In all, we constructed 16 matches: four each for the four comparisons outlined in Section~\ref{sec:multiple_comparison_groups}.
For each comparison, we built a separate match using our four estimates of the propensity score: the MLE, the $\ell_{1}$-regularized estimate, the posterior mean with respect to a standard prior, and the estimate from BART. 
In the process of creating these matches, we found that some individuals had some missing covariate values.
We dealt with this by augmenting our covariate set with a binary indicator for missingness and imputing missing values with the average value; please see \citet{Rosenbaum1984} and \citet{Rosenbaum2008} for a longer discussion on how to deal with missingness in matching. 
We further noticed that there were some instances in which all subjects who were missing values of a certain covariate had the same treatment or control status.
We dropped these individuals from our analysis.

\begin{figure}[H]
\centering
\includegraphics[width = \textwidth]{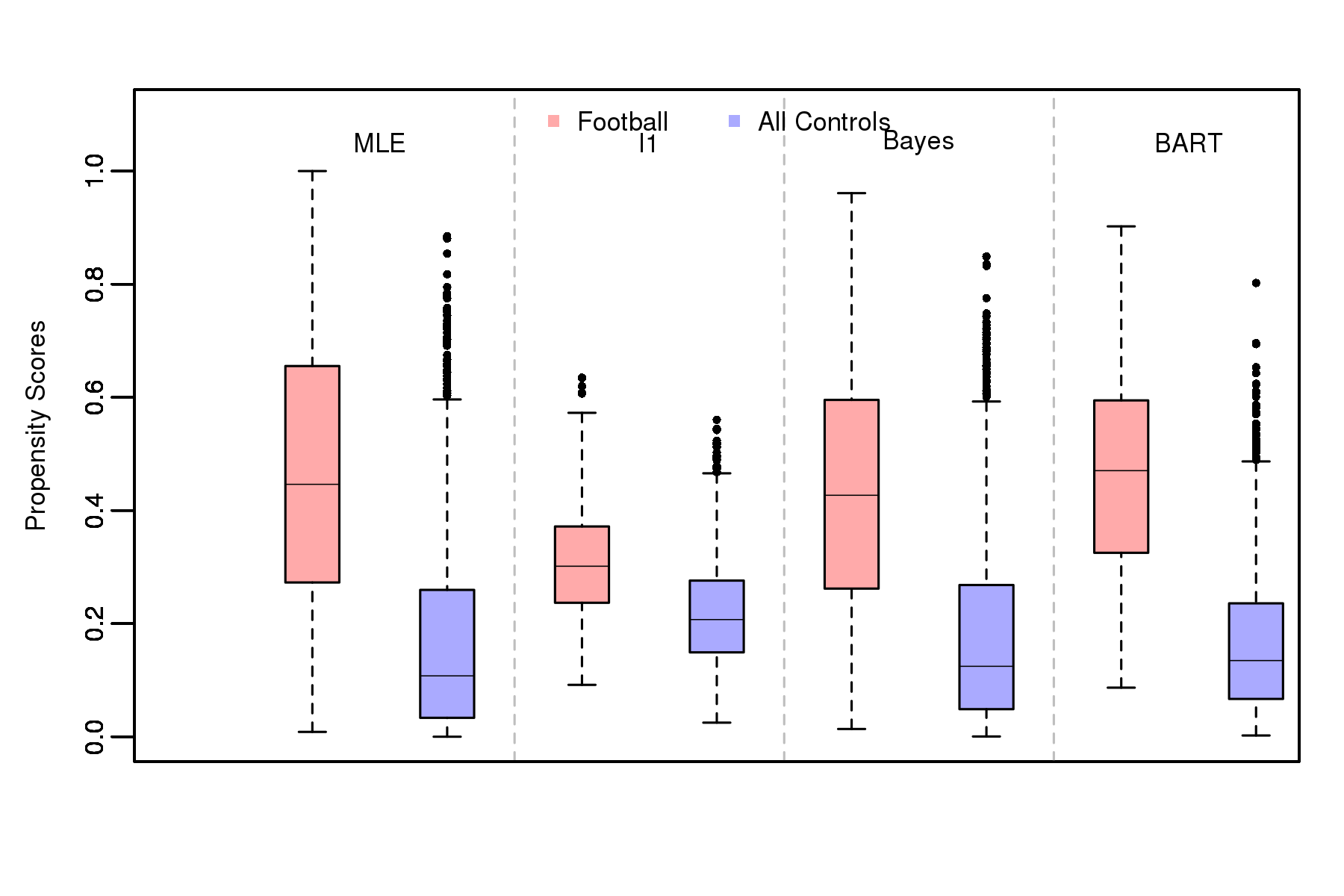}
\caption{Comparative box plots of the estimated propensity scores for both treated (red) and control subjects (blue)}
\label{fig:propensity1}
\end{figure}

Figure~\ref{fig:propensity1} shows comparative box plots of the distribution of the four estimated propensity scores for treated and control subjects for Match 1 between football players and all controls. 
Corresponding figures for the other matches are available in Appendix~\ref{app:additional_figures}.
We observe that the majority of treated subjects tend to have larger estimated propensity scores than the control subjects.
However, there are also some treated subjects with lower propensity scores than all control subjects.
There are also some control subjects with larger propensity scores than all treated subjects.
These are examples of individuals that \textit{lack common support}; essentially for these individuals, there are no subjects with the opposite treatment status but similar propensity score and covariates.
As such, it is generally not feasible to get reliable estimates of the causal effect of interest from them.
We drop these individuals from our analysis before running the variable ratio matching procedure.
See \citet{Hill2013} for a more involved discussion and strategies for assessing the lack of common support in observational studies.

Table~\ref{tab:match_details} summarizes the number of subjects dropped for lack of common support and the number of variables for which we did not achieve adequate balance for each match considered. 
As a reminder, Comparison 1 is between football players and all control subjects, Comparison 2 is between football players and sport-playing controls, Comparison 3 is between football players and non-sport controls, and Comparison 4 is between the sport-playing and non-sport controls.
In Comparison 4, we take ``playing a non-collision'' sport as the treatment.

\begin{table}[H]
\caption{Summary of the matches constructed. $n_{miss}$ reports the number of subjects dropped because their treatment assignment was perfectly predicted by the fact that they were missing values of a particular covariate. $n_{cs}$ counts the number of subjects dropped because their propensity scores indicated a lack of common support. For both $n_{miss}$ and $n_{cs}$, the numbers within the parentheses are the  number of treated and control subjects dropped, respectively. Similarly $n_{total}$ reports the number of subjects included in the match. For each of $n_{miss}$, $n_{cs},$ and $n_{total}$, we report the number of treated and control subjects within parentheses dropped or included, respectively. We also report the number of covariates for which the standardized difference after matching exceed 0.2 in absolute value}
\label{tab:match_details}
\centering
\begin{tabular}{l|c|cccc} \hline
~ & $n_{miss}$ & Propensity &  $n_{cs}$ & $n_{total}$ & Number Imbalanced \\ \hline
\multirow{4}{*}{Comparison 1} & \multirow{4}{*}{32 (0 / 32))} & MLE & 203 (26 / 177) & 1481 (447 / 1034)  & 0 \\
&  & $\ell_{1}$ & 106 (11 / 95) & 1549 (492 / 1057) & 1 \\
&  & Bayes & 141 (15 / 126) & 1476 (442 / 1034) & 0 \\ 
&  & BART & 525 (15 / 520) & 911 (323 / 588) & 0 \\ \hline
\multirow{4}{*}{Comparison 2} & \multirow{4}{*}{9 (0 / 9)} & MLE & 92 (44 / 48) & 671 (289 / 382) & 0 \\
&  & $\ell_{1}$ & 10 (10 / 0) & 992 (496 / 496) & 4 \\
&  & Bayes & 30 (11 / 19) & 734 (319 / 415) & 0 \\ 
&  & BART & 207 (87 / 120) & 628 (304 / 324) & 2 \\ \hline
\multirow{4}{*}{Comparison 3} & \multirow{4}{*}{25 (2 / 23)} & MLE & 249 (47 / 202) & 869 (290 / 579) & 3 \\
&  & $\ell_{1}$ & 85 (18 / 67) & 1047 (437 / 610) & 0 \\
&  & Bayes & 162 (48 / 114) & 872 (312 / 560) & 1 \\ 
&  & BART & 330 (71 / 259) & 639 (260 / 379) & 1 \\ \hline
\multirow{4}{*}{Comparison 4} & \multirow{4}{*}{11 (8 / 3)} & MLE & 0 (0 / 0) & 1487 (483 / 1004) & 18\\
&  & $\ell_{1}$ & 22 (2 / 20) & 1297 (556 / 741) & 0 \\
&  & Bayes & 50 (11 / 39) & 1091 (420 / 671) & 1 \\ 
&  & BART & 258 (55 / 208) & 773 (323 / 450) & 1 \\ \hline
\end{tabular}
\end{table}

As mentioned above, for each comparison, we will pick the match that achieves adequate balance on the baseline covariates and drops the fewest number of subjects.
In particular, we will use the match based on the MLE of the propensity score for the comparison between football players and all controls (Comparison 1), the match based on the Bayesian estimate of the propensity score for the comparison between football players and sport-playing controls (Comparison 2), and the matches based on the $\ell_{1}$-regularized estimates of the propensity scores for the comparisons between football players and non-sport-playing controls (Comparison 3) and between the two sub-groups of controls (Comparison 4). 

Interestingly, we found that the BART propensity score estimates tended to over-fit the data slightly.
In particular, the estimated propensity scores for treated subjects tended to be much larger than those for the control subjects.
This resulted in many more subjects being dropped, out of consideration for lack of common support.  

Table~\ref{tab:matched_set_composition} in Appendix~\ref{app:additional_figures} reports the composition of the matched sets for each comparisons.

Recall that the goal of our matching procedure is to balance the distribution of several potential confounders between treatment and control group.
Since we cannot hope to achieve exact equality in these distributions, we instead aim to make the standardized difference in covariate means between the treated and control groups as small as possible.
Our objective in this work is to achieve standardized differences that are less than 0.2 in absolute value, since regression-based covariate adjustment has been shown to remove biases due to covariate imbalances of this magnitude \citep{Cochran1973, Silber2001}. 
Tables~\ref{tab:balance_match1_health} and ~\ref{tab:balance_match1_substance} summarize the balance of several covariates related to health, substance abuse, personality, and life plan that were used in match selected for Comparison 1. 

We note that many of these covariates are ordinal.
In building our matches, we treated these as continuous variables and note that the post-match standardized differences were all less than 0.2 in absolute value.
We further treat them as ordinal in our covariate adjustment. 
Nevertheless, for ease of presentation, in Tables~\ref{tab:balance_match1_health} and~\ref{tab:balance_match1_substance}, we treat these covariates as categorical, reporting the percentage of subjects who fall within each category.
In doing so, we should point out that there is no general guarantee that adequate balance on the continuous scale implies adequate balance on the discrete scale (or vice-versa).
Nevertheless, we find it encouraging that for nearly all of the variables in question, we do have adequate balance on both scales.

In general, football players and all controls appear adequately balanced on most of these covariates prior to matching.
Nevertheless, matching generally tended to decrease these standardized differences. 
That being said, the two groups were still imbalanced along several of these covariates prior to matching.
For instance, prior to matching, football players were, on average, 10 pounds heavier than all controls (160.08 vs 150.01) and were more likely to identify as black or African-American (27.45\% vs 17.72\%), exercise more than five times a week (32.25\% vs 23.63\%), and more likely to have high desire to go to college (73.7\% vs 62.35\%). 
Additionally, they were less likely to never experience muscle or joint aches (11.13\% vs 20.17\%) or be regular smokers (14.01\% vs 20.47\%). 
The standardized differences along these covariates all exceed 0.2 prior to matching.
After matching, however, all of these standardized differences were less than 0.2. 

\begin{table}[H]
\centering
\caption{Comparison of average baseline variables related to demographics, general health, and life plan before and after matching football players to all controls. Before matching, control values are unweighted. After matching, control values are weighted according to the composition of the matched set (see Table~\ref{tab:matched_set_composition}). Standardized differences greater than 0.2 are bolded.}
\label{tab:balance_match1_health}
\tiny
\begin{tabular}{lcccccc} \hline
~ & \multicolumn{2}{c}{Before Matching} & \multicolumn{2}{c}{After Matching} & \multicolumn{2}{c}{Standardized Differences} \\
 ~ & Football & All Controls & Football & All Controls & Before Matching & After Matching \\ \hline \\
Age in 2008 (yrs) & 28.76 & 29.06 &28.71 &28.71 & -0.173 & 0 \\
Weight in 1994-95 (lbs) & 160.08 & 150.01 & 157.7 & 155.31 & \bf{0.263} & 0.063 \\
Height in 1994-95 (in) & 68.49 & 68.48 & 68.36 & 68.26 & 0.002 &0.024 \\
Subject's reported race & ~ & ~ & ~ & ~ & ~ & ~ \\
\hspace{2em}White (\%) & 66.03  & 72.55  & 67.79 & 66.34  & -0.175 & 0.039 \\
\hspace{2em} Black (\%) & 27.45  & 17.72 & 25.5 & 26.75  & \bf{0.282} & -0.036 \\
\hspace{2em} Native American (\%) & 2.69 & 2.63 & 2.46  & 2.32 & 0.005 & 0.011 \\
\hspace{2em} Asian (\%) & 2.69  & 4.83  & 2.46  & 2.81  & -0.155 &-0.025 \\
\hspace{2em} Other (\%) & 4.99  & 6.26 & 4.92 & 5.34 &-0.072 & -0.024 \\
Subject's assessment of his general health & ~ & ~ & ~ & ~ & ~ & ~ \\
\hspace{2em}Excellent (\%)&39.35 & 30.91& 38.7 & 37.59 & \bf{0.22} & 0.029 \\
\hspace{2em}Very good (\%) & 39.54 & 39.56 & 39.82 & 40.79  & 0 & -0.025 \\
\hspace{2em}Good (\%) & 16.89 & 24.58  & 17  & 17.57 & \bf{-0.25} & -0.019 \\
\hspace{2em}Fair (\%) & 3.84 & 4.42  & 4.03  & 3.74  & -0.037 & 0.018 \\ 
\hspace{2em}Poor(\%) & 0.38  & 0.48 & 0.45  & 0.31  & -0.019 & 0.028 \\
How often past week subject exercised & ~ & ~ & ~ & ~ & ~ & ~ \\
\hspace{2em}Not at all (\%) &15.55  & 23.21 & 14.99  & 15.82  & \bf{-0.257} & -0.028 \\
\hspace{2em}1 - 2 times (\%) & 28.21 & 30.79 & 29.31  & 28.99  & -0.072 & 0.009 \\
\hspace{2em}3 - 4 times (\%) & 23.99  & 22.32 & 25.28  & 26.68  &0.05 &-0.041 \\
\hspace{2em}5+ times (\%) & 32.25 & 23.63  & 30.43  & 28.52 & \bf{0.236} &0.052 \\
How often subject had & ~ & ~ & ~ & ~ & ~ & ~ \\
\hspace{1em} Headaches & ~ & ~ & ~ & ~ & ~ & ~ \\
\hspace{2em}Never (\%) &10.17 &11.52  &10.07  & 11.5  & -0.055 & -0.059 \\
\hspace{2em}A few times (\%) & 70.44  & 67.18 & 70.25  & 66.23 & 0.089 & 0.11 \\
\hspace{2em}Once a week (\%) & 16.51  & 17.6  & 17  & 20.54  & -0.037 & -0.119 \\
\hspace{2em}Almost every day (\%) & 2.5 & 2.98 & 2.24  & 1.64 & -0.039 & 0.047 \\
\hspace{2em}Every day (\%) & 0.38 & 0.66 & 0.45 & 0.1  & -0.052 & 0.067 \\
\hspace{1em} Unexplained physical weakness & ~ & ~ & ~ & ~ & ~ & ~ \\
\hspace{2em}Never (\%) & 61.23 & 60.44  & 61.3  & 63.41  & 0.02 & -0.054 \\
\hspace{2em}A few times (\%) & 34.74  & 33.59  & 34.68  & 33.26  & 0.03 & 0.037 \\
\hspace{2em}Once a week (\%) & 3.84  & 4.65  & 3.8  & 2.63  & -0.052 & 0.075 \\
\hspace{2em}Almost every day (\%) & 0.19  & 0.95  & 0.22  & 0.66  & -0.162 & -0.092 \\
\hspace{2em}Every day (\%) & 0  & 0.3  & 0  & 0.04  & -0.15 & -0.022 \\
\hspace{1em} Dizziness & ~ & ~ & ~ & ~ & ~ & ~ \\
\hspace{2em}Never (\%) & 69.1  & 65.21 & 68.9 & 68.93  & 0.105 & -0.001 \\
\hspace{2em}A few times (\%) & 25.53  & 29.47  & 27.07 & 28.26 & -0.113 & -0.034 \\
\hspace{2em}Once a week (\%) & 4.22 & 3.64  & 3.58  & 2.09 & 0.037 & 0.094 \\
\hspace{2em}Almost every day (\%) & 0.77 & 1.01 & 0.45 & 0.46& -0.034 & -0.002 \\
\hspace{2em}Every day (\%) & 0.38 & 0.6  & 0  & 0.25  & -0.041 & -0.049 \\
\hspace{1em} Muscle or joint aches & ~ & ~ & ~ & ~ & ~ & ~ \\
\hspace{2em}Never (\%) & 11.13 & 20.17  & 10.96 & 13.57& \bf{-0.34} & -0.098 \\
\hspace{2em}A few times (\%) & 50.86 & 53.34  & 52.35  & 53.37  & -0.062 & -0.026 \\
\hspace{2em}Once a week (\%) & 28.6 & 20.41  & 28.86 & 24.26  & \bf{0.233} & 0.131 \\
\hspace{2em}Almost every day (\%) & 7.29& 5.01  & 6.94  & 7.61& 0.114 & -0.034 \\
\hspace{2em}Every day (\%) & 2.11  & 1.01  & 0.89  & 1.19  & 0.102 & -0.028 \\
\hspace{1em} Trouble sleeping& ~ & ~ & ~ & ~ & ~ & ~ \\ 
\hspace{2em}Never (\%) & 44.15 & 42.96 & 44.3 & 44.13 & 0.03 & 0.004 \\
\hspace{2em}A few times (\%)&37.62 &34.67&38.93 &37.83 &0.077 & 0.029 \\
\hspace{2em}Once a week (\%)&11.32 &12.83 &11.41 &11.35 &-0.059 & 0.002 \\
\hspace{2em}Almost every day (\%)	&4.61 &7.04 &3.8 &5.63 )&-0.139&-0.104 \\
\hspace{2em}Every day (\%) & 2.3 & 2.45 & 1.57 & 1.06 &-0.012&0.042 \\
Seriously considered suicide in past year (\%) & 9.4 & 10.2 ) & 7.38 & 6.49  & -0.034 & 0.038  \\ 
Desire to go to college & ~ & ~ & ~ & ~ & ~ & ~ \\
\hspace{1em}1 - low (\%)&3.26 &5.43 &2.68 &	2.5 &-0.144&0.012\\
\hspace{1em}2 (\%)&1.34&4.18 &1.34&2.3 )&-0.259&-0.087\\
\hspace{1em}3 - medium (\%)&7.49&12.17 &7.38 &8.09 &\bf{-0.212}&-0.032\\
\hspace{1em}4 (\%)&14.01 &15.63 &	14.77& 14.42 & -0.058 & 0.012\\
\hspace{1em}5 - high (\%)&73.7 &62.35 & 73.6 & 72.59&\bf{0.317} &0.028\\
Likelihood of going to college & ~ & ~ & ~ & ~ & ~ & ~ \\
\hspace{1em}1 - low (\%)	&3.84 &6.38 &3.13 &2.98&-0.157&0.009\\
\hspace{1em}2 (\%)	&4.03 &6.38 &4.47&4.12 &-0.143&	0.021\\
\hspace{1em}3 - medium (\%)&	14.2&16.17 &13.2&13.29 &-0.07&-0.003\\
\hspace{1em}4 (\%)	&23.99 &	23.57 &23.94 &	24.76&0.013& -0.024\\
\hspace{1em}5 - high (\%)&53.74 &47.2 &55.03 &54.74 &0.165& 0.007 \\ \hline
\end{tabular}

\end{table}

\begin{table}[H]
\centering
\caption{Comparison of average baseline variables related to substance abuse, personality, and self-perception before and after matching football players to all controls.Before matching, control values are unweighted. After matching, control values are weighted according to the composition of the matched set (see Table~\ref{tab:matched_set_composition}). Standardized differences greater than 0.2 are bolded. Some subjects were not asked every questions and these variables are marked $^{\dagger}.$}
\label{tab:balance_match1_substance}
\tiny
\begin{tabular}{lcccccc} \hline
~ & \multicolumn{2}{c}{Before Matching} & \multicolumn{2}{c}{After Matching} & \multicolumn{2}{c}{Standardized Differences} \\
 ~ & Football & All Controls & Football & All Controls & Before Matching & After Matching \\ \hline \\
Ever tried cigarette smoking (\%) & 53.55  & 56.09 & 51.23 & 49.65  & -0.064 & 0.04 \\
Smoked regularly (\%) & 14.01 & 20.47& 12.08& 12 & \bf{-0.226} & 0.003 \\
Ever drank alcohol (\%) & 55.09 & 55.01 & 53.02  & 49.72 & 0.002 & 0.083 \\
How often subject drank alcohol in past year & ~ & ~ & ~ & ~ & ~ & ~ \\
\hspace{2em}Almost every day (\%)	&0.77 &1.19 &0.45 &1.43 &-0.058&-0.134\\
\hspace{2em}3 - 5 days/week (\%)&	2.88 &3.28 &2.24 &2.48  &-0.03&-0.018\\
\hspace{2em}1 - 2 days/week (\%)&	6.33 &6.86 &5.82 &5.52 & -0.027&0.015\\
\hspace{2em}2 - 3 days/month (\%)	&10.17 &7.1 &10.29 &7.4 &0.132&0.124\\
\hspace{2em}Once a month or less (\%)&	10.56 &11.1 &10.07 &9.17 &-0.022&0.036\\
\hspace{2em}Once or twice in past year (\%)&	14.4 &14.98&14.09 &14.41&-0.021&-0.011\\
\hspace{2em}Never (\%)	&9.79 &10.14 &	10.07 &9.31 &-0.015&0.032\\
\hspace{2em}Not applicable$^{\dagger}$ (\%)&45.11 &45.35 &46.98 &50.28 &-0.005&-0.066\\
How often subject had 5+ drinks in a row & ~ & ~ & ~ & ~ & ~ & ~ \\
\hspace{2em}Almost every day (\%)&0.77 &0.72 &0.45 &0.25 &0.008&0.029\\
\hspace{2em}3 - 5 days/week (\%)&	1.73 &2.39 &1.12 &2&-0.061&-0.082\\
\hspace{2em}1 - 2 days/week (\%)&	4.99 &4.95&4.92&4.61 &0.002&0.018\\
\hspace{2em}2 - 3 days/month (\%)&	6.91 &5.31 &6.94 &4.92&0.081&0.102\\
\hspace{2em}Once a month or less (\%)&	6.14 &5.61 &5.82&5.87  & 0.028&-0.003\\
\hspace{2em}Once or twice in past year (\%)&	7.29&8.41&6.94 & 6.84 &-0.053&0.004\\
\hspace{2em}Never (\%)&	16.89 &17.06&16.33&15.93 &-0.006&0.014\\
\hspace{2em} Not applicable$^{\dagger}$ (\%)&55.28&55.55 &57.49 &59.58&	-0.005&-0.042 \\ 
How much subject agree that: & ~ & ~ & ~ & ~ & ~ & ~\\
\hspace{1em} You never get sad & ~ & ~ & ~ & ~ & ~ & ~ \\
\hspace{2em}Strongly agree (\%)&4.03 &5.37 &4.03 &5.66 &-0.083&-0.101\\
\hspace{2em}Agree (\%)&15.16 &13.9 &15.21 &15.12 &0.045&0.003 \\
\hspace{2em}Neither agree nor disagree (\%)&19.77 & 20.88&19.02 &18.35 &-0.035&0.021\\
\hspace{2em}Disagree (\%)&53.17 &49.7 &54.14 &50.88 &0.087&0.082 \\
\hspace{2em}Strongly disagree (\%)&7.87 &10.08 &7.61 &9.98 &-0.101&-0.108 \\
\hspace{1em} You are physically fit & ~ & ~ & ~ & ~ & ~ & ~ \\
\hspace{2em}Strongly agree (\%)&42.61 &31.38 &40.94 &43.1 &\bf{0.289}&-0.056 \\
\hspace{2em}Agree (\%)&47.6&47.49 &48.77 &45.28 &0.003&0.088 \\
\hspace{2em}Neither agree nor disagree (\%)&8.06 &14.26&8.28 & 9.02 &\bf{-0.269}&-0.032 \\
\hspace{2em}Disagree (\%)&1.73 & 6.03 & 2.01 & 2.51 &\bf{-0.34} & -0.039 \\
\hspace{2em}Strongly disagree (\%) &0& 0.72 &0 &0.07&\bf{-0.233} & -0.024 \\
\hspace{1em} You feel socially accepted & ~ & ~ & ~ & ~ & ~ & ~ \\
\hspace{2em}Strongly agree (\%) &38.2 &31.8 & 36.24 & 38.89& 0.167 & -0.069 \\
\hspace{2em}Agree (\%)&54.13 &54.47 &56.15 &51.99&-0.009&0.105 \\
\hspace{2em}Neither agree nor disagree (\%)&6.14 &9.61 &6.26 &7.13 &-0.173&-0.043 \\
\hspace{2em}Disagree (\%)&1.15 &3.76 &0.89 &1.76 &\bf{-0.255}&-0.084 \\
\hspace{2em}Strongly disagree (\%)&0.38 &0.24&0.45 &0.24 &0.031&0.044 \\
How much subject feels parents care about him & ~ & ~ & ~ & ~ & ~ & ~ \\
\hspace{2em}Not at all (\%) & 0.19 & 0.36 & 0 & 0.11  & -0.044 & -0.03 \\
\hspace{2em}Very little (\%) & 0.77  & 0.6  & 0.89 & 0.6 &0.025 &0.043 \\
\hspace{2em}Somewhat (\%) &2.3 & 2.63 & 1.57 & 2.15 & -0.027 & -0.048\\
\hspace{2em}Quite a bit (\%) &10.56 & 12.95 & 9.62 &9.96 & -0.096 &-0.014 \\
\hspace{2em}Very much (\%) & 85.99& 83.17 & 87.7 & 86.94  & 0.1 & 0.027\\
\hspace{2em}Missing (\%) & 0.19 &0.18& 0.22  & 0.24 & 0.003 & -0.003\\ \hline

\end{tabular}
\end{table}

\section{Discussion}
\label{sec:discussion}

In this protocol, we have proposed a matched observational study on the effect of playing football in adolescence on mental health in early adulthood using data from the Add Health study.
A key strength of the proposed study is the use of prospectively collected longitudinal data from a nationally-representative sample. 
Another strength is our ability to control and adjust for a wide range of potential confounders through the combined use of propensity score matching and covariance adjustment. 
The use of multiple control groups enables us to probe one source of potential unmeasured confounding, namely the possibility that controls who played a non-collision sport may differ systematically from controls who did not play any sport at all. 
Carrying out a sensitivity analysis also enables us to assess the degree to which our results may change in the presence of unmeasured confounders.


There are also several limitations of the proposed study.
Perhaps most glaringly, is that we were unable to measure directly subjects' exposure to football-related head trauma.
The Add Health dataset did not record whether the subjects actually played football or not; instead, it simply asked whether or not they participated in or intended to participate in school football.
In this way, our treatment group likely contains some subjects who intended to play football but ended up not playing for some reason.
Additionally, the dataset lacked a direct measure of subjects' exposure to head trauma (football-related or otherwise), the age of first exposure to football, or any further information about football participation like position played.
Recent research suggests that there may be a thresholded dose-response relationship between later-life impairment and the cumulative number of head impacts \citep{Montenigro2017} and that there may be an association between age of first exposure to American football and later-life impairment \citep{Alosco2017}.
Since nearly 25\% of eligible subjects were missing the primary outcome of interest, the CES-D score measured in 2008, our results may also be limited by selection bias.
Concern over this is mitigated slightly, however, by the fact  that our attrition analysis found that playing football was not significantly associated with a higher or lower likelihood of missing the primary outcome.
Further, in this work we were primarily interested in inference about a single fixed treatment effect. 
It is probably more realistic to think that the effect of playing football is quite heterogeneous. 
In light of the potential heterogeneity, it may be more natural to seek inference about the average treatment effect.
Recently, \citet{Fogarty2016} considered this problem in the context of paired observational studies and developed procedures for sensitivity analysis.
Extending his framework to our setting, in which our matched sets contain a variable number of controls, may be of separate methodological interest. 
Finally, we must point out that results of our study may not generalize exactly to current middle- and high-school football players due to changes in playing style, training technique, and rules aimed at improving safety. 

\newpage
\bibliography{addhealth_refs}

\newpage
\appendix

\section{Baseline Covariates}
\label{app:baseline_covariates}

Below is a full list of baseline covariates used in our study. Also indicated are the Add Health variable names.

\textbf{Demographic Data}
\begin{itemize}
\item{Age}
\item{Height}
\item{Weight}
\item{Self-identified race (H1GI6A -- H1GI6E). \textit{Note: subjects could identify with multiple racial backgrounds}}
\item{Single best description of racial background (H1GI8)}
\item{Race, as determined by Wave I interviewer (H1GI9)}
\item{Whether subject was born in the United States (H1GI11)}
\item{Whether subject lived in a rural, suburban, or urban environment (H1IR12)}
\end{itemize}

\textbf{Family Background}
\begin{itemize}
\item{How far mother / father went in school (H1RM1 / H1RF1)}
\item{Whether mother / father was born in the United States (H1RM2 / H1RF2)}
\item{Type of work done by mother / father (H1RM4 / H1RF4)}
\item{Whether mother / father works for pay (H1RM5 / H1RF5)}
\item{Whether mother / father worked for pay at any time in past 12 months (H1RM6 / H1RF6)} 
\item{Whether mother / father received public assistance such as welfare (H1RM9 / H1RF9)}
\item{Has mother / father ever smoked cigarettes (H1RM14 / H1RF14)}
\item{How close subject felt to mother / father (H1WP9 / H1WP13)}
\item{How much subject thinks mother / father cares about him (H1WP10 / H1WP14)}
\item{Whether the subject felt that his mother / father was warm and loving toward him, most of the time (H1PF1 / H1PF23)}
\item{Whether subject's mother encouraged him to be independent (H1PF2)}
\item{Whether subject's mother provides constructive feedback when subject has done something wrong (H1PF3)}
\item{Whether subject is satisfied with the way he communicates with mother / father (H1PF4 / H1PF24)}
\item{Whether subject is overall satisified with his relationship with his mother / father (H1PF5 / H1PF25)}
\end{itemize}

\textbf{School Performance and Experience}
\begin{itemize}
\item{Whether subject was presently in school or was in school when interview was conducted (H1GI18)}
\item{Current or most recent grade in school (H1GI20)}
\item{Grade in English / mathematics/ history or social sciences / science at most recent grading period (H1ED11 -- H1ED14)}
\item{Frequency with which subject had trouble getting along with his teachers (H1ED15)}
\item{Frequency with which subject had trouble paying attention in school (H1ED16)}
\item{Frequency with which subject had trouble completing homework assignments (H1ED17)}
\end{itemize}

\textbf{General Health}
\begin{itemize}
\item{Number of times times subject exercised in the past week (H1DA6)}
\item{Self-report of general health (H1GH1)}
\item{How often in previous year subject had each of the following conditions:
\begin{itemize}
\item{Headache (H1GH2)}
\item{Feeling hot all over suddenly (H1GH3)}
\item{Stomach ache or upset stomach (H1GH4)}
\item{Cold sweats (H1GH5)}
\item{Physical weakness (H1GH6)}
\item{Sore throat or cough (H1GH7)}
\item{Tiredness (H1GH8)}
\item{Painful or very frequent urination (H1GH9)}
\item{Feeling very sick (H1GH10)}
\item{Waking up feeling tired (H1GH11)}
\item{Skin problems such as itching or pimples (H1GH12)}
\item{Dizziness (H1GH13)}
\item{Chest pains (H1GH14)}
\item{Aches, pains, or soreness in muscles or joints (H1GH15)}
\item{Poor appetite (H1GH17)}
\item{Trouble falling or staying asleep (H1GH18)}
\item{Trouble relaxing (H1GH19)}
\end{itemize}
}
\item{Number of times in past year has subject had a routine physical exam / had psychological or emotional counseling / attended drug or alcohol abuse treatment program (H1HS1 / H1HS3 / H1HS5)}
\end{itemize}

\textbf{Propensity for ``risky behavior''}
\begin{itemize}
\item{Frequency with which subject wore wear a helmet when riding a bicycle / motorcycle (H1GH39 / H1GH41)}
\item{Frequency with which subject rode a motorcycle (H1GH40)}
\item{Frequency with which subject wore seatbelt while riding in or driving a car (H1GH42)}
\item{Frequency with which subject drove after drinking alcohol in the past month (H1GH43)}
\end{itemize}

\textbf{Propensity for delinquency behavior:} We created a single measure of delinquency by counting the number of below activities in which the subject engaged at least once in the previous year.  
\begin{itemize}
\item{Paint graffiti or signs (H1DS1)}
\item{Deliberately damaged property that did not belong to him (H1DS2)}
\item{Lie to parents or guardians (H1DS3)}
\item{Take something from a store without paying for it (H1DS4)}
\item{Get into a serious physical fight (H1DS5)}
\item{Hurt someone badly enough to need bandages (H1DS6)}
\item{Run away from home (H1DS7)}
\item{Drive a car without owner's permission (H1DS8)}
\item{Stole something worth $> \$50$ (H1DS9)}
\item{Went into a building to steal something (H1DS10)}
\item{Use or threaten to use a weapon to get something from someone (H1DS11)}
\item{Sell marijuana or other drugs (H1DS12)}
\item{Stole something worth $< \$50$ (H1DS13)}
\item{Took part in a large group fight (H1DS14)}
\item{Were loud, rowdy, or unruly in public place (H1DS15)}
\end{itemize}

\textbf{Substance Abuse}
\begin{itemize}
\item{Whether subject has ever tried cigarette smoking (H1TO1)}
\item{Age when subject smoked first whole cigarette (H1TO2)}
\item{Whether subject has ever smoked regularly (1+ cigarettes/day for 30 days) (H1TO3)}
\item{Age at which subject began to smoke regularly (H1TO4)}
\item{How many days did subject smoke (H1TO5)}
\item{How many cigarettes did subject smoke each day, during the past 30 days (H1TO7)}
\item{Has the subject tried to quit smoking in the past six months (H1TO8)}
\item{Number of days on which subject has chewed tobacco in the past month (H1TO10)}
\item{Age when subject first start chewing tobacco (H1TO11)}
\item{Whether the subject has drunk alcohol more than two or three times (H1TO12)}
\item{Whether the subject drank alcohol without parents or family (H1TO13)}
\item{Age when subject first drank alcohol without family (H1TO14)}
\item{Number of days in the past year when subject drank (H1TO15)}
\item{Number of drinks did subject have each time they drank (H1TO16)}
\item{Number of times over the past year that subject had more than five drinks in a row (H1TO17)}
\item{Number of times in the past year has subject gotten very drunk (H1TO18)}
\item{Age when subject first tried marijuana (H1TO30)}
\item{Number of times subject has used marijuana in his life (H1TO31)}
\item{Number of times subject has used marijuana in the past month (H1TO32)}
\item{Age when subject first tried cocaine (H1TO34)}
\item{Number of times subject has used cocaine in his life (H1TO35)}
\item{Number of times subject has sued cocaine in past month (H1TO36)}
\item{Age when subject first tried inhalants (H1TO37)}
\item{Number of times subject has used inhalants in his life (H1TO38)}
\item{Number of times subject has used inhalants in past month (H1TO39)}
\item{Age when subject first tried any other illegal drug (H1TO40)}
\item{Number of times subject used any other illegal drug in his life (H1TO41)}
\item{Number of times subject used any other illegal drug in past month (H1TO42)}
\end{itemize}

\textbf{Exposure to Violence}: Subjects were asked how often 
\begin{itemize}
\item{Saw someone shot or stabbed in past year(H1FV1)}
\item{Pulled a knife or gun on subject in past year (H1FV2)}
\item{Someone shot the subject in past year(H1FV3)}
\item{Someone cut or stabbed subject in the past year (H1FV4)}
\item{Subject got into a physical fight in the past year (H1FV5)}
\item{Subject was jumped in the past year (H1FV6)}
\item{Subject pulled a knife or gun on someone in the past year (H1FV7)}
\item{Subject shot or stabbed someone in the past year (H1FV8)}
\item{Subject carried a weapon to school (H1FV9)}
\end{itemize}

\textbf{Suicidality}
\begin{itemize}
\item{Whether subject ever seriously thought about committing suicide in the past year (H1SU1)}
\item{How many times did subject actually attempt suicide in the past year (H1SU2)}
\item{Whether any of subject's friends had attempted suicide in the past year (H1SU4)}
\item{Whether any subject's family members had attempted suicide in the past year (H1SU6)}
\end{itemize}

\textbf{``Protective Factors''}: Each subject was asked how much they felt that
\begin{itemize}
\item{Adults care about him (H1PR1)}
\item{Teachers care about him (H1PR2)}
\item{His parents care about him (H1PR3)}
\item{His friends care about him (H1PR4)}
\item{People in his family understand him (H1PR5)}
\item{He wants to leave home (H1PR6)}
\item{He and his family have fun together (H1PR7)}
\item{His family pays attention to him (H1PR8)}
\end{itemize}

\textbf{Feelings}: Each subject was asked how often each of the following statements were true during the past week
\begin{itemize}
\item{He bothered by things that usually don't bother him(H1FS1)}
\item{He did not feel like eating or his appetite was poor. (H1FS2)}
\item{He felt that he could not shake off the blues, even with help from his family and friends (H1FS3)}
\item{He felt that he was just as good as other people (H1FS4)}
\item{He had trouble keeping his mind on what he was doing (H1FS5)}
\item{He felt depressed (H1FS6)}
\item{He felt that you were too tired to do things (H1FS7)}
\item{He felt hopeful about the future (H1FS8)}
\item{He thought his life had been a failure (H1FS9)}
\item{He felt fearful (H1FS10)}
\item{He were happy (H1FS11)}
\item{He talked less than usual (H1FS12)}
\item{He felt lonely (H1FS13)}
\item{People were unfriendly to him (H1FS14)}
\item{He enjoyed life (H1FS15)}
\item{He felt sad (H1FS16)}
\item{He felt that people disliked you (H1FS17)}
\item{It was hard to get started doing things (H1FS18)}
\item{He felt life was not worth living (H1FS19)}
\end{itemize}

\textbf{Personality}: Each subject was asked how much they agreed with each of the following statements
\begin{itemize}
\item{He never argues with anyone (H1PF7)}
\item{When he gets what he wants, it's usually because he worked hard for it (H1PF8)}
\item{He never gets sad (H1PF10)}
\item{He never criticizes other people (H1PF13)}
\item{He usually goes out of his way to avoid having to deal with problems in his life (H1PF14)}
\item{Difficult problems make him very upset (H1PF15)}
\item{When making decisions, he usually goes with his "gut feeling" without thinking too much about the consequences of each alternative (H1PF16)}
\item{When he has a problem to solve, one of the first things he does is get as many facts about the problem as possible (H1PF18)}
\item{When he is attempting to find a solution to a problem, he usually tries to think of as many different ways to approach the problem as possible (H1PF19)}
\item{When making decisions, he generally uses a systematic method for judging and comparing alternatives (H1PF20)}
\item{After carrying out a solution to a problem, he usually tries to analyze what went right and what went wrong (H1PF21)}
\item{He has a lot of energy (H1PF26)}
\item{He seldom gets sick (H1PF27)}
\item{When he does get sick, he gets better quickly (H1PF28)}
\item{He is well coordinated (H1PF29)}
\item{He has a lot of good qualities (H1PF30)}
\item{He is physically fit (H1PF31)}
\item{He has a lot to be proud of (H1PF32)}
\item{He likes himself just the way he is (H1PF33)}
\item{He feels like he is doing everything just about right (H1PF34)}
\item{He feels socially accepted (H1PF35)}
\item{He feels loved and wanted (H1PF36)}
\end{itemize}

\textbf{Expectations, Employment, and Income}
\begin{itemize}
\item{How much do subject wants to go to college (H1EE1)}
\item{How likely is it that subject will go to college (H1EE2)}
\item{How many hours does subject spend working for pay in typical non-summer week (H1EE4)}
\end{itemize}

\textbf{Variables from the in-school questionnaire}
\begin{itemize}
\item{How often subject smoked / drank / get drunk in the past year (S59A / S59B / S59C)}
\item{How often subject did something dangerous on a dare in the past year (S59E)}
\item{How often subject lied to parents or guardians in the past year (S59F)}
\item{How often subject skipped school without an excuse in the past year (S59G)}
\item{How many times in a normal week does subject work, play, or exercise hard enough to sweat and breathe heavily  (S63)}

Each subject was asked how much he agreed with the following statements
\item{He feels close to people at his school (S62B)}
\item{He feels like he is part of his school (S62C)}
\item{He feels like the students at his school are prejudiced (S62G)}
\item{He is happy to be at his school (S62I)}
\item{The teachers at his school treat him fairly (S62L)}
\item{He feels safe in my neighborhood (S62Q)}
\item{He feels safe in my school (S62R)}

\end{itemize}

\section{Additional Figures and Tables}
\label{app:additional_figures}

\begin{figure}[H]
\centering
\includegraphics[width = \textwidth]{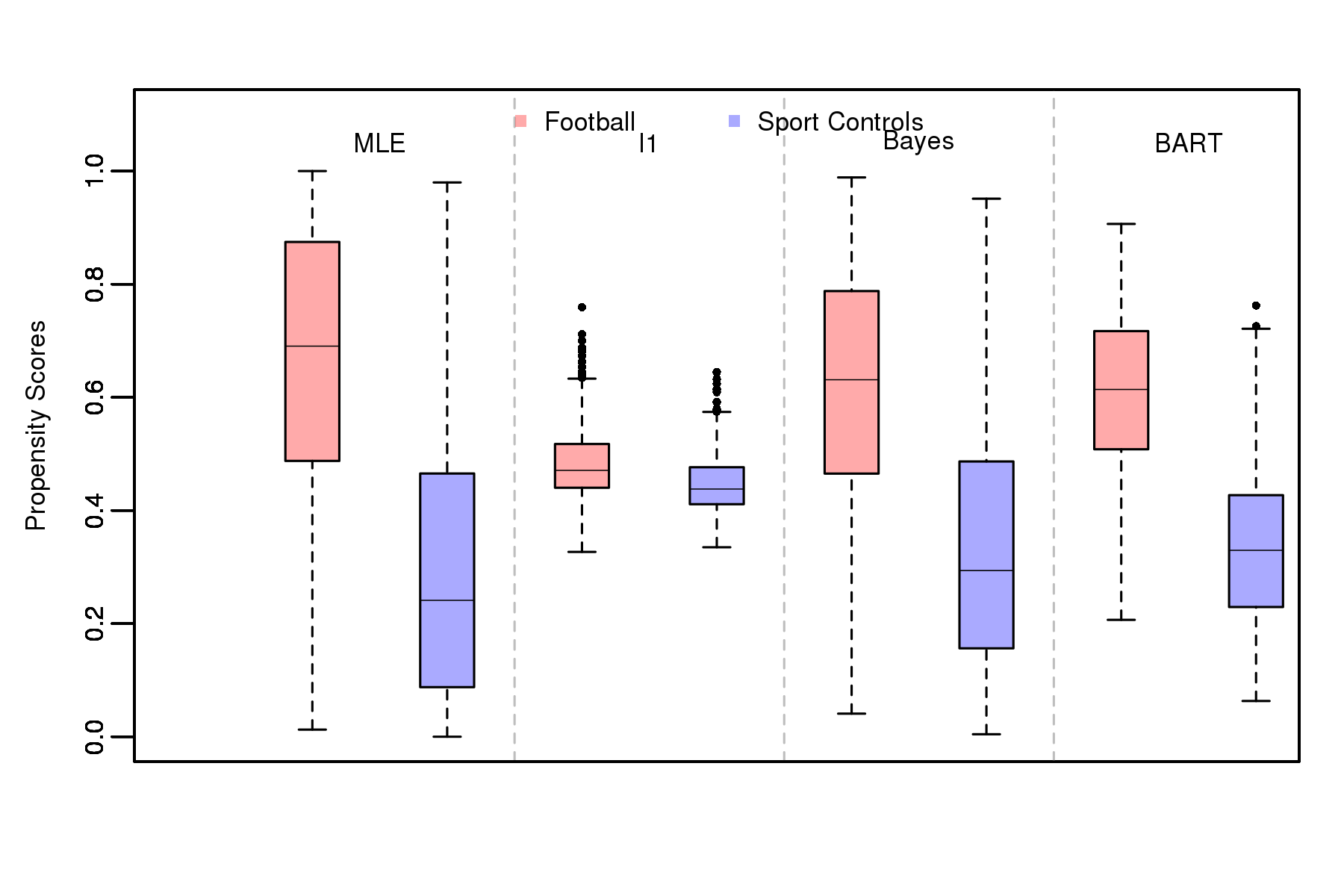}
\caption{Analog of Figure~\ref{fig:propensity1} for Comparison 2, between football players (red) and sport-playing controls (blue).} 
\label{fig:propensity2}
\end{figure}

\begin{figure}[H]
\centering
\includegraphics[width = \textwidth]{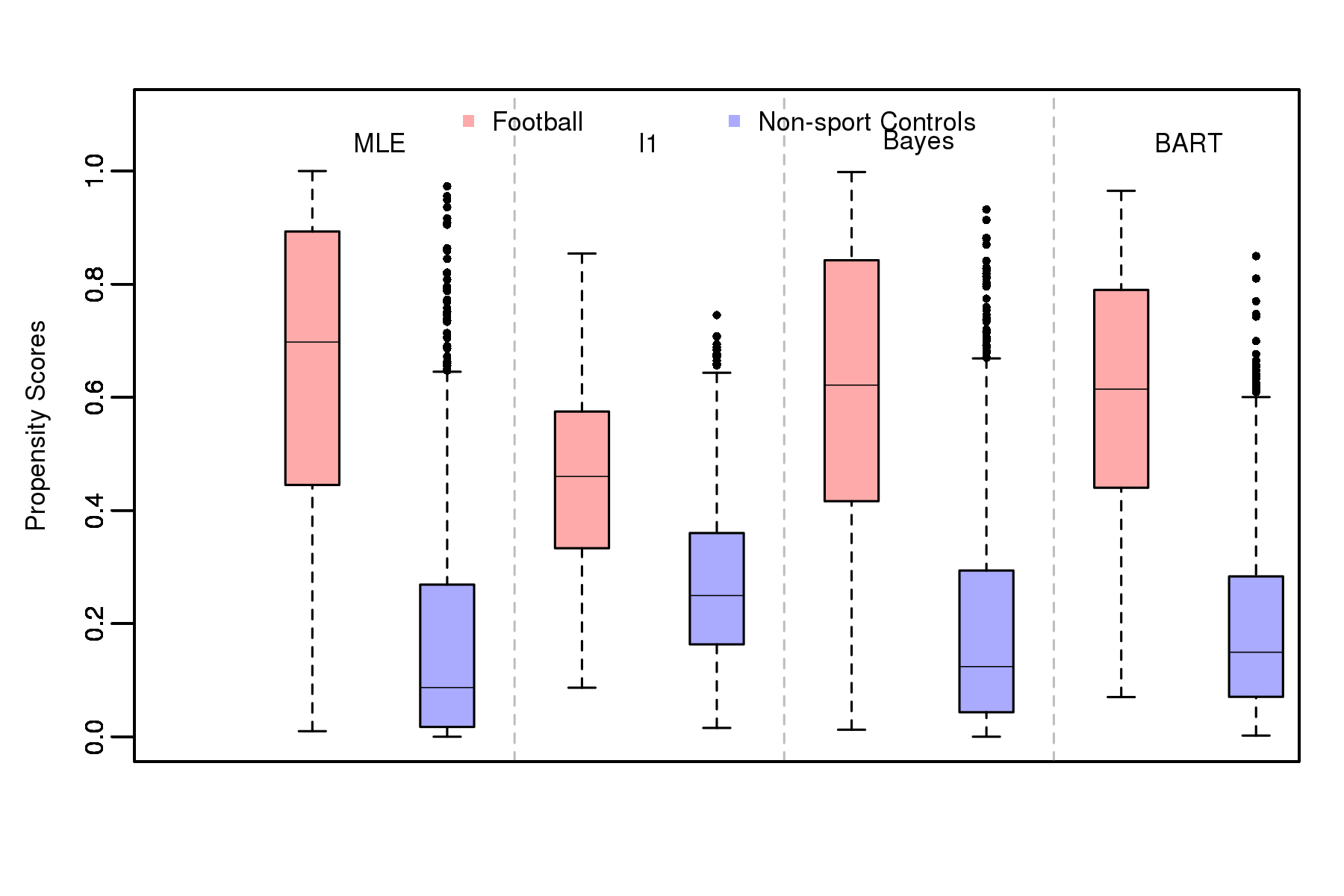}
\caption{Analog of Figure~\ref{fig:propensity1} for Comparison 3, between football players (red) and non-sport-playing controls (blue).}
\label{fig:propensity3}
\end{figure}

\begin{figure}[H]
\centering
\includegraphics[width = \textwidth]{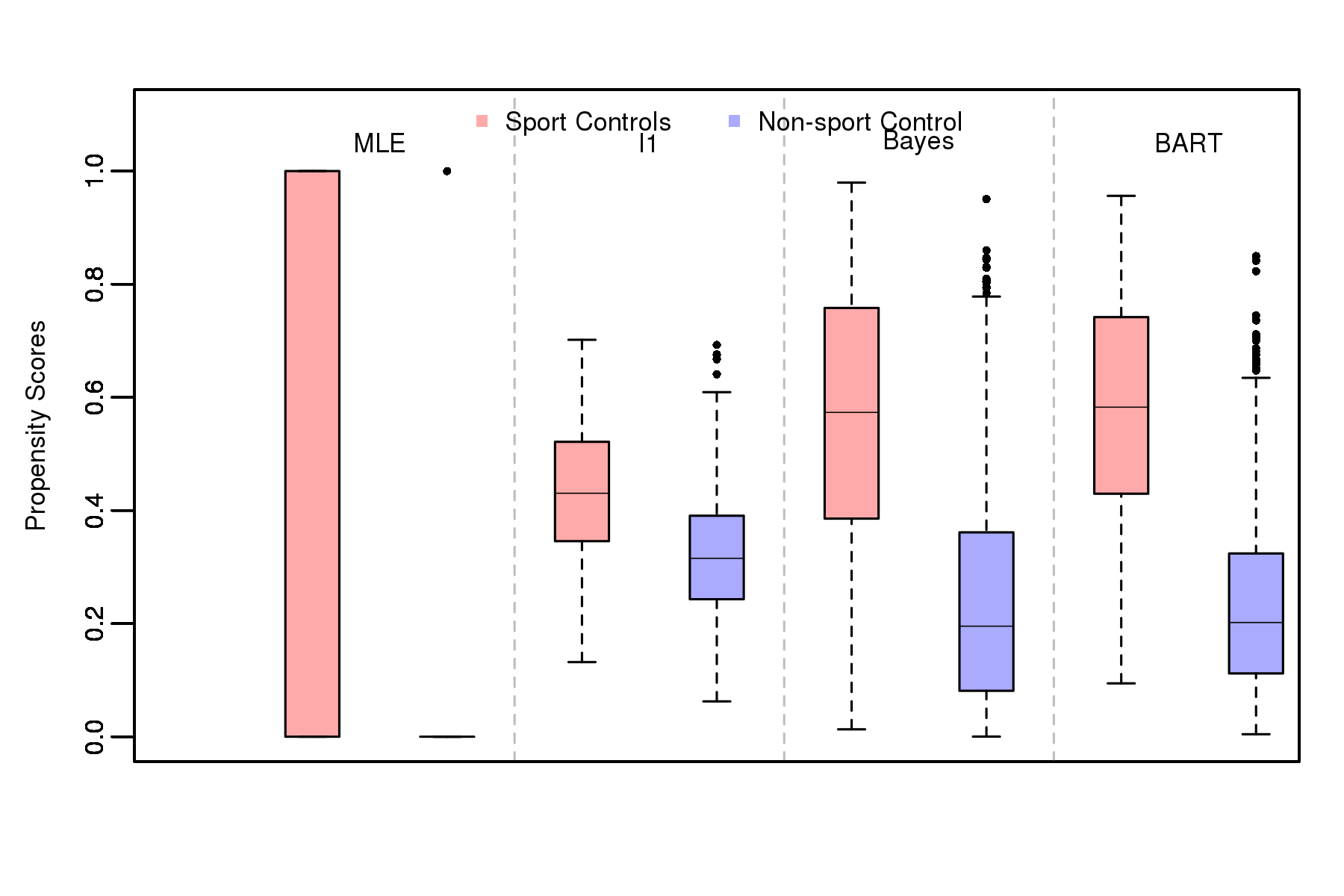}
\caption{Analog of Figure~\ref{fig:propensity4} for Comparison 4, between sport-playing controls (red) and non-sport-playing controls (blue)}
\label{fig:propensity4}
\end{figure}

Figure~\ref{fig:propensity2},~\ref{fig:propensity3}, and~\ref{fig:propensity4} are analogs of Figure~\ref{fig:propensity1} for Comparisons 2 -- 4.

Table~\ref{tab:matched_set_composition} shows the composition of the constructed matches.

\begin{table}[H]
\centering
\caption{Composition of the matched sets constructed for each comparison. The first number in the Composition column is the number of exposed subjects (i.e. football players in Comparison 1 -- 3 and sport controls in Comparison 4) and the second number is the number of controls.}
\label{tab:matched_set_composition}
\begin{tabular}{ccccc} \hline
Composition & Comparison 1 & Comparison 2 & Comparison 3 & Comparison 4 \\ \hline
1:1 & 290 & 264 & 342 & 426 \\
1:2 & 47 & 31 & 45 & 87 \\
1:3 & 49 & 16 & 23 & 33 \\
1:4 & 9 & 1 & 20 & 9 \\
1:5 & 17 & 5 & 0 & 0 \\
1:6 & 10 & 2 & 3 & 1 \\
1:7 & 8 & 0 & 0 & 0 \\
1:8 & 2 & 0 & 1 & 0 \\
1:9 & 3 & 0 & 0 & 0 \\
1:10 & 1 & 0 & 0 & 0 \\
1:11 & 2 & 0 & 0 & 0 \\
1:12 & 0 & 0 & 0 & 0 \\
1:13 & 0 & 0 & 0 & 0 \\
1:14 & 0 & 0 & 0 & 0 \\
1:15 & 17 & 0 & 0 & 0 \\ \hline
\end{tabular}
\end{table}

Tables~\ref{tab:balance_match2_health} --\ref{tab:balance_match4_substance} are analogs of Tables~\ref{tab:balance_match1_health} and~\ref{tab:balance_match1_substance} for Comparisons 2 -- 4.

\begin{table}[H]
\centering
\caption{Comparison of average baseline variables related to demographics, general health, and life plan before and after matching football players to sport controls. Before matching, control values are unweighted. After matching, control values are weighted according to the composition of the matched set (see Table~\ref{tab:matched_set_composition}). Standardized differences greater than 0.2 are bolded.}
\label{tab:balance_match2_health}
\tiny
\begin{tabular}{lcccccc} \hline
~ & \multicolumn{2}{c}{Before Matching} & \multicolumn{2}{c}{After Matching} & \multicolumn{2}{c}{Standardized Differences} \\
 ~ & Football & Sport Controls & Football & Sport Controls & Before Matching & After Matching \\ \hline \\
Age in 2008 (yrs)& 28.76 & 28.94 & 28.66 & 28.66 & -0.109 & 0 \\
Weight in 1994-95 (lbs)& 160.08 & 149.66 & 155.52 & 151.88 & \bf{0.29} & 0.101 \\
Height in 1994-95 (in)& 68.49 & 69.13 & 68.41 & 68.56 & -0.157 & -0.038 \\
Subject's reported race & ~ & ~ & ~ & ~ & ~ & ~ \\
\hspace{2em}White (\%)& 66.03 & 71.64 & 68.34 & 67.99 & -0.138 & 0.009 \\
\hspace{2em}Black (\%)& 27.45 & 21.97 & 27.27 & 25.77 & 0.144 & 0.04 \\
\hspace{2em}Native American (\%)& 2.69 & 1.8 & 2.19 & 2.31 & 0.067 & -0.009 \\
\hspace{2em}Asian (\%)& 2.69 & 5.74 & 2.19 & 4.1 & -0.185 & -0.115 \\
\hspace{2em}Other (\%)& 4.99 & 3.44 & 2.51 & 3.87 & 0.086 & -0.076 \\
Subject's assessment of his general health & ~ & ~ & ~ & ~ & ~ & ~ \\
\hspace{2em}Excellent (\%)&  39.35 & 37.05 & 41.07 & 38.15 & 0.054 & 0.069 \\
\hspace{2em}ery good (\%)& 39.54 & 42.3 & 40.44 & 45.05 & -0.064 & -0.108 \\
\hspace{2em}Good (\%)& 16.89 & 17.21 & 15.36 & 14.03 & -0.01 & 0.041 \\
\hspace{2em}Fair (\%) & 3.84 & 2.95 & 2.82 & 2.46 & 0.055 & 0.023 \\
\hspace{2em}Poor(\%)&	0.38 & 0.33 & 0.31 & 0.3 & 0.011 & 0 \\
How often past week subject exercised & ~ & ~ & ~ & ~ & ~ & ~ \\
\hspace{2em}Not at all (\%)& 15.55 & 20.33 & 13.17 & 17.36 & -0.146 & -0.128 \\
\hspace{2em}1 - 2 times (\%)& 28.21 & 28.52 & 29.15 & 27.78 & -0.008 & 0.035 \\
\hspace{2em}3 - 4 times (\%)& 23.99 & 21.48 & 26.96 & 22.58 & 0.069 & 0.119 \\
\hspace{2em}5+ times (\%)&	32.25 & 29.51& 30.72 & 32.27 & 0.068 & -0.038 \\
How often subject had & ~ & ~ & ~ & ~ & ~ & ~ \\
\hspace{1em} Headaches & ~ & ~ & ~ & ~ & ~ & ~ \\
\hspace{2em}Never (\%)& 10.17 & 11.64 & 10.97 & 12.79 & -0.055 & -0.068 \\
\hspace{2em}A few times (\%)& 70.44 & 65.9 & 70.85 & 63.56 & 0.113 & 0.181 \\
\hspace{2em}Once a week (\%)& 16.51 & 20.16 & 15.99 & 22.03 & -0.11 & -0.181 \\
\hspace{2em}Almost every day (\%)& 2.5 & 1.8 & 1.88 & 1.31 & 0.053 & 0.044 \\
\hspace{2em} Every day (\%)& 0.38 & 0.33 & 0.31 & 0.31 & 0.011 & 0 \\
\hspace{1em} Unexplained physical weakness & ~ & ~ & ~ & ~ & ~ & ~ \\
\hspace{2em}Never (\%)& 61.23 & 62.62 & 63.32 & 62.08 & -0.033 & 0.029 \\
\hspace{2em}A few times (\%)& 34.74 & 31.8 & 33.86 &	33.92 & 0.071 & -0.002 \\
\hspace{2em}Once a week (\%)& 3.84 & 4.43 & 2.82 & 2.95 & -0.034 & -0.008 \\
\hspace{2em}Almost every day (\%)& 0.19 & 0.49& 0 & 0.94 & -0.063 & -0.199 \\
\hspace{2em}Every day (\%)& 0 & 0.49 & 0 &	0.1 & -0.138 & -0.029 \\
\hspace{1em} Dizziness & ~ & ~ & ~ & ~ & ~ & ~ \\
\hspace{2em}Never (\%)& 69.1 & 64.92 & 66.77 & 68.75 & 0.103 & -0.048 \\
\hspace{2em}A few times (\%)& 25.53 &	30.98 & 29.47 & 28.92 & -0.141& 0.014 \\
\hspace{2em}Once a week (\%)& 4.22 & 2.62 & 3.76 & 1.39 & 0.098 & 0.144 \\
\hspace{2em}Almost every day (\%)& 0.77  & 0.98 & 0& 0.63 & -0.027 & -0.079 \\
\hspace{2em}Every day (\%)& 0.38 & 0.33 & 0 & 0.31 & 0.011 & -0.06 \\
\hspace{1em} Muscle or joint aches & ~ & ~ & ~ & ~ & ~ & ~ \\
\hspace{2em}Never (\%)& 11.13 & 15.57 & 11.6 & 14.02 & -0.154 & -0.084 \\
\hspace{2em}A few times (\%)& 50.86 & 53.44 & 51.72 & 52.98 & -0.059 & -0.029 \\
\hspace{2em}Once a week (\%)& 28.6 & 21.8 & 30.72 & 22.24 & 0.178 & 0.222 \\
\hspace{2em}Almost every day (\%)& 7.29 & 7.87 & 5.64 & 9.82 & -0.025 & -0.182 \\
\hspace{2em}Every day (\%)& 2.11 & 1.15 & 0.31 & 0.94 & 0.084 & -0.054\\
\hspace{1em} Trouble sleeping& ~ & ~ & ~ & ~ & ~ & ~ \\ 
\hspace{2em}Never (\%)& 44.15 & 44.26 & 45.77 & 45.44 & -0.003 & 0.007 \\
\hspace{2em}A few times (\%)& 37.62 & 35.74 & 38.87 & 36.82 & 0.045 & 0.049 \\
\hspace{2em}Once a week (\%)& 11.32 & 11.97  &11.29 & 10.59 & -0.023 & 0.025 \\
\hspace{2em}Almost every day (\%)& 4.61 & 6.72 & 2.82 & 5.99 & -0.108 & -0.162 \\
\hspace{2em}Every day (\%)& 2.3 & 1.15 & 1.25 & 1.15 & 0.097 & 0.009\\
Seriously considered suicide in past year (\%)& 9.4 & 7.87 & 4.7 & 7.34 & 0.062 & -0.106  \\ 
Desire to go to college & ~ & ~ & ~ & ~ & ~ & ~ \\
\hspace{2em}1 - low (\%)& 3.26 & 2.95 & 2.82 & 2.35 & 0.02 & 0.031 \\
\hspace{2em}2 (\%)& 1.34 & 2.95 & 1.57 & 2.35 & -0.135 & -0.066 \\
\hspace{2em}3 - medium (\%)& 7.49 & 9.67 & 5.96 & 7.86 & -0.091 & -0.079 \\
\hspace{2em}4 (\%)& 14.01 & 11.64 & 14.42 & 11.19 & 0.08 & 0.11 \\
\hspace{2em}5 - high (\%)& 73.7 & 72.46 & 75.24 & 76.25 & 0.032 & -0.026 \\
Likelihood of going to college & ~ & ~ & ~ & ~ & ~ & ~ \\
\hspace{2em}1 - low (\%)& 3.84 & 4.1 & 2.19 & 2.4 & -0.015 & -0.012 \\
\hspace{2em}2 (\%)& 4.03 & 4.43 & 3.76 & 4.25 & -0.023 & -0.028 \\
\hspace{2em}3 - medium (\%) & 14.2 & 11.31 & 12.54 & 9.81 & 0.098 & 0.093 \\
\hspace{2em}4 (\%)& 23.93 & 25.41 & 25.08 & 26.04 & -0.038 & -0.026 \\
\hspace{2em}5 - high (\%)& 53.74 & 54.26 & 56.43 & 57.49 & -0.012 & -0.025\\ \hline
\end{tabular}

\end{table}

\begin{table}[H]
\centering
\caption{Comparison of average baseline variables related to substance abuse, personality, and self-perception before and after matching football players (FB) to sport controls (SC). Before matching, sport control values are unweighted. After matching, sport control values are weighted according to the composition of the matched set (see Table~\ref{tab:matched_set_composition}). Standardized differences greater than 0.2 are bolded.}
\label{tab:balance_match2_substance}
\tiny
\begin{tabular}{lcccccc} \hline
~ & \multicolumn{2}{c}{Before Matching} & \multicolumn{2}{c}{After Matching} & \multicolumn{2}{c}{Standardized Differences} \\
 ~ & FB & SC & FB & SC & Before Matching & After Matching \\ \hline \\
Ever tried cigarette smoking (\%)& 53.55 & 53.28 & 46.08 & 48.21 & 0.006 & -0.049 \\
Smoked regularly (\%)& 14.01 & 14.26 & 7.52 & 11.24 (49/415)& -0.008 & -0.123 \\
Ever drank alcohol (\%)& 55.09 & 54.26 & 44.83 & 49.95 & 0.019 & -0.118 \\
How often subject drank alcohol in past year & ~ & ~ & ~ & ~ & ~ & ~ \\
\hspace{2em}Almost every day (\%)& 0.77 & 0.82 & 0.31 & 0.63 & -0.007 & -0.041 \\
\hspace{2em}3 - 5 days/week (\%)& 2.88 & 3.61 & 0.63 & 3.12 & -0.048 & -0.164 \\
\hspace{2em}1 - 2 days/week (\%)& 6.33 & 6.39 & 4.7 & 5.24 & -0.003 & -0.025 \\
\hspace{2em}2 - 3 days/month (\%)& 10.17 &	6.56 & 8.78 & 7.6 & 0.146 & 0.047 \\
\hspace{2em}Once a month or less (\%)& 10.56 & 11.97 & 8.15 & 10.66 & -0.052 & -0.092 \\
\hspace{2em}Once or twice in past year (\%)& 14.4 & 14.1 & 11.91 & 12.94 & 0.01 & -0.034 \\
\hspace{2em}Never (\%)& 9.79 & 10.49 & 10.03 & 9.67 & -0.027 & 0.014 \\
\hspace{2em}Not applicable (\%)& 45.11 & 46.07 & 55.49 & 50.15 & -0.019 & 0.107 \\
How often subject had 5+ drinks in a row & ~ & ~ & ~ & ~ & ~ & ~ \\
\hspace{2em}Almost every day (\%)& 0.77 & 0.49 & 0 & 0 & 0.039 & 0 \\
\hspace{2em}3 - 5 days/week (\%)& 1.73 & 2.46 & 0.63 & 2.47 & -0.06 & -0.152 \\
\hspace{2em}1 - 2 days/week (\%)& 4.99 & 5.41 & 3.13 & 4.99 & -0.022 & -0.096 \\
\hspace{2em}2 - 3 days/month (\%)& 6.91 & 3.61 & 5.64 & 2.8 & 0.163 & 0.14 \\
\hspace{2em}Once a month or less (\%)& 6.14 & 6.23 & 3.76 & 5.55 & -0.004 & -0.085 \\
\hspace{2em}Once or twice in past year (\%)&7.29 &9.51 &5.96 &	8.87 & -0.093 & -0.123 \\
\hspace{2em}Never (\%)& 16.89 & 15.57 & 15.36 & 15.19 & 0.041 & 0.005 \\
\hspace{2em} Not applicable (\%)& 55.28 & 56.72 & 65.52 & 60.14 & -0.03 & 0.111 \\
How much subject agree that: & ~ & ~ & ~ & ~ & ~ & ~\\
\hspace{1em} You never get sad & ~ & ~ & ~ & ~ & ~ & ~ \\
\hspace{2em}Strongly agree (\%)& 4.03 & 4.92 & 3.45 & 4.68 & -0.05 & -0.069 \\
\hspace{2em}Agree (\%)& 15.16 & 15.25 & 15.99 & 16.17 & -0.003 & -0.006 \\
\hspace{2em}Neither agree nor disagree (\%) & 19.77 & 19.51 & 18.5 & 19.25 & 0.008 & -0.022 \\
\hspace{2em}Disagree (\%)&	53.17 & 51.31 & 55.8 & 51.44 & 0.043 & 0.1 \\
\hspace{2em}Strongly disagree (\%)& 7.87 & 8.85 & 6.27 & 8.46 & -0.041 & -0.092 \\
\hspace{1em} You are physically fit & ~ & ~ & ~ & ~ & ~ & ~ \\
\hspace{2em}Strongly agree (\%) & 42.61 & 42.62 & 40.75 & 45.66 & 0 & -0.114 \\
\hspace{2em}Agree (\%)& 47.6 & 45.41 & 50.47 & 43.83 & 0.05 & 0.153 \\
\hspace{2em}Neither agree nor disagree (\%)& 8.06 & 9.02 & 6.9 & 8.73 & -0.04 & -0.076 \\
\hspace{2em}Disagree (\%)&	1.73 & 2.46 & 1.88 & 1.47 & -0.06 & 0.034 \\
\hspace{2em}Strongly disagree (\%)& 0 & 0.33 & 0 & 0.31 & -0.112 & -0.107 \\
\hspace{1em} You feel socially accepted & ~ & ~ & ~ & ~ & ~ & ~ \\
\hspace{2em}Strongly agree (\%)&	38.2 & 36.72 & 37.93 & 39 & 0.035 & -0.025 \\
\hspace{2em}Agree (\%)& 54.13 & 54.26 & 56.11 & 52.77 & -0.003 & 0.077 \\
\hspace{2em}Neither agree nor disagree (\%)& 6.14 & 6.72 & 5.96 & 6.82 & -0.027 & -0.041 \\
\hspace{2em}Disagree (\%)&	1.15 & 1.97 & 0 &1.41 & -0.079 & -0.136 \\
\hspace{2em}Strongly disagree (\%)& 0.38 & 0 & 0 & 0 & 0.088 & 0 \\
How much subject feels parents care about him & ~ & ~ & ~ & ~ & ~ & ~ \\
\hspace{2em}Not at all (\%)&	0.19 & 0.49 & 0.31 & 0.31 & -0.063 & 0 \\
\hspace{2em}Very little (\%)&	0.77 & 0.49 & 0.31 & 0.63 & 0.039 & -0.044 \\
\hspace{2em}Somewhat (\%)& 2.3 & 1.8 & 1.25 & 2.22 & 0.04 & -0.077 \\
\hspace{2em}Quite a bit (\%)& 10.56 & 11.48 & 10.97 & 10.85 & -0.034 & 0.004 \\
\hspace{2em}Very much (\%)& 85.99 & 85.41 & 87.15 & 85.99 & 0.019 & 0.038 \\ \hline

\end{tabular}
\end{table}

\begin{table}[H]
\centering
\caption{Comparison of average baseline variables related to demographics, general health, and life plan before and after matching football players to non-sport controls. Before matching, control values are unweighted. After matching, control values are weighted according to the composition of the matched set (see Table~\ref{tab:matched_set_composition}). Standardized differences greater than 0.2 are bolded.}
\label{tab:balance_match3_health}
\tiny
\begin{tabular}{lcccccc} \hline
~ & \multicolumn{2}{c}{Before Matching} & \multicolumn{2}{c}{After Matching} & \multicolumn{2}{c}{Standardized Differences} \\
 ~ & FB & NSC & FB & NSC & Before Matching & After Matching \\ \hline \\
Age in 2008 (yrs) & 28.76 & 29.13 & 28.6&28.63&-0.207& -0.015 \\
Weight in 1994-95 (lbs)& 160.08 150.21 & 155.1 & 152.44 & \bf{0.25} & 0.067 \\
Height in 1994-95 (in)& 68.49 & 68.11 & 68.11 & 67.97 & 0.092 & 0.034 \\
Subject's reported race & ~ & ~ & ~ & ~ & ~ & ~ \\
\hspace{2em}White (\%)& 66.03& 73.08 & 66.59 & 69.78 & -0.181 & -0.082 \\
\hspace{2em}Black (\%)& 27.45& 15.29 & 26.96 & 23.72 &\bf{0.343} & 0.091 \\
\hspace{2em}Native American (\%)& 2.69 & 3.1 & 2.07 & 2.08 & -0.03 & -0.001 \\
\hspace{2em}Asian (\%)& 2.69 & 4.32 & 2.3 & 2.32 & -0.111 & -0.001 \\
\hspace{2em}Other (\%)& 4.99 & 7.88 & 5.07 & 5.7 & -0.147 & -0.032 \\
Subject's assessment of his general health & ~ & ~ & ~ & ~ & ~ & ~ \\
\hspace{2em}Excellent (\%)&	39.35 &27.39 & 38.02 & 35.82 & \bf{0.299} & 0.055 \\
\hspace{2em}Very good (\%)& 39.54 & 37.99 & 41.71 & 38.02 & 0.038 & 0.09 \\
\hspace{2em}Good (\%)&16.89&28.8& 15.9 & 21.48 & \bf{-0.356} & -0.167 \\
\hspace{2em}Fair (\%)&	3.84 & 5.25 & 3.92 &	 4.09 & -0.084 & -0.01 \\
\hspace{2em}Poor(\%)&	0.38 & 0.56 & 0.46 & 0.6 & -0.032 & -0.024 \\
How often past week subject exercised & ~ & ~ & ~ & ~ & ~ & ~ \\
\hspace{2em}Not at all (\%)&	15.55 & 24.86 &15.67&19.58&	\bf{-0.29} &  -0.121 \\
\hspace{2em}1 - 2 times (\%)& 28.21 & 32.08 & 29.26 &	28.25 & -0.101 & 0.026 \\
\hspace{2em}3 - 4 times (\%)& 23.99 &22.8 & 25.81 & 27.77 & 0.034 & -0.055 \\
\hspace{2em}5+ times (\%)&	32.25 & 20.26 & 29.26 & 24.4 & \bf{0.318} & 0.129 \\
How often subject had & ~ & ~ & ~ & ~ & ~ & ~ \\
\hspace{1em} Headaches & ~ & ~ & ~ & ~ & ~ & ~ \\
\hspace{2em}Never (\%)& 10.17 & 11.44 & 11.06 & 11.05 & -0.049 & 0 \\
\hspace{2em}A few times (\%)& 70.44 & 67.92 & 70.74 & 70.9 & 0.066 & -0.004 \\
\hspace{2em}Once a week (\%)& 16.51 & 16.14 & 15.9 & 14.8 & 0.012 0.035 \\
\hspace{2em}Almost every day (\%)& 2.5 & 3.66 & 1.84 & 2.57 & -0.084 & -0.053 \\
\hspace{2em}Every day (\%)& 0.38 & 0.84 & 0.46 & 0.67 & -0.077 & -0.035 \\
\hspace{1em} Unexplained physical weakness & ~ & ~ & ~ & ~ & ~ & ~ \\
\hspace{2em}Never (\%)& 61.23 & 59.19 & 62.23 & 65.75 & 0.05 & -0.087 \\
\hspace{2em}A few times (\%)& 34.74 & 34.62 & 34.1 & 30.12 & 0.003 & 0.1 \\
\hspace{2em}Once a week (\%)& 3.84 & 4.78 & 3.46 & 3.8 &  -0.057 & -0.021 \\
\hspace{2em}Almost every day (\%)& 0.19 & 1.22 & 0.23 & 0.33 & -0.176 & -0.016 \\
\hspace{2em}Every day (\%)& 0 & 0.19 & 0 & 0 & -0.096 & 0 \\
\hspace{1em} Dizziness & ~ & ~ & ~ & ~ & ~ & ~ \\
\hspace{2em}Never (\%)& 69.1 & 65.38 & 70.05 & 70.1 & 0.095 & -0.001 \\
\hspace{2em}A few times (\%)& 25.53 & 28.61 & 26.73 & 25.41 & -0.084 & 0.036 \\
\hspace{2em}Once a week (\%)& 4.22 & 4.22 & 3 & 3.45 & 0 & -0.027 \\
\hspace{2em}Almost every day (\%)& 0.77 & 1.03 & 0.23 & 1.04 & -0.034 & -0.105 \\
\hspace{2em}Every day (\%)& 0.38 & 0.75 &	0 & 0 & -0.063 & 0 \\
\hspace{1em} Muscle or joint aches & ~ & ~ & ~ & ~ & ~ & ~ \\
\hspace{2em}Never (\%)& 11.13 & 22.8 & 11.98 & 14.78 & -0.4 & -0.096 \\
\hspace{2em}A few times (\%)& 50.86 & 53.28 &	52.07 & 55.05 & -0.058 & -0.071 \\
\hspace{2em}Once a week (\%)&	28.6 & 19.6 & 28.57 & 24.36 & \bf{0.246} & 0.115 \\
\hspace{2em}Almost every day (\%)& 7.29 & 3.38 & 6.68 & 4.38 & 0.195 & 0.115 \\
\hspace{2em}Every day (\%)& 2.11 & 0.94 & 0.69 & 1.44 & 0.106 & -0.068\\
\hspace{1em} Trouble sleeping& ~ & ~ & ~ & ~ & ~ & ~ \\ 
\hspace{2em}Never (\%)& 44.15 & 42.21 & 44.93 & 45.14 & 0.047 & -0.005 \\
\hspace{2em}A few times (\%)& 37.62 & 34.05 & 37.79 & 36.42 & 0.088 & 0.034 \\
\hspace{2em}Once a week (\%)& 11.32 & 13.32 & 11.75 & 11.66 & -0.074 & 0.003 \\
\hspace{2em}Almost every day (\%)& 4.61 & 7.22 & 3.92 & 5.01 & -0.139 & -0.058 \\
\hspace{2em}Every day (\%)& 2.3 &  3.19 & 1.61 & 1.77 & -0.067 & -0.012 \\
Seriously considered suicide in past year (\%)& 9.4 (49/521)& 11.54 & 7.14 & 6.11 & -0.085 & 0.041   \\ 
Desire to go to college & ~ & ~ & ~ & ~ & ~ & ~ \\
\hspace{2em}1 - low (\%)& 3.26  & 6.85 & 2.76 & 2.36 & \bf{-0.212} & 0.024 \\
\hspace{2em}2 (\%)& 1.34 & 4.88 & 1.15 & 2.74 & \bf{-0.279} & -0.125 \\
\hspace{2em}3 - medium (\%)& 7.49 & 13.6 & 6.91 & 10.63 & \bf{-0.253} & -0.154 \\
\hspace{2em}4 (\%)& 14.01 & 17.92 & 14.52 & 17.73 & -0.13 & -0.107 \\
\hspace{2em}5 - high (\%)& 73.7 & 56.57 & 74.65 & 66.47 & \bf{0.448} & \bf{0.214} \\
Likelihood of going to college & ~ & ~ & ~ & ~ & ~ & ~ \\
\hspace{2em}1 - low (\%)& 3.84 & 7.69 & 3.46 & 3.61 & \bf{-0.213} & -0.008 \\
\hspace{2em}2 (\%)& 4.03 & 7.5 & 3.92 & 3.98 & -0.19 & -0.004 \\
\hspace{2em}3 - medium (\%)& 14.2 & 18.95 & 13.36 & 16.32 & -0.156 & -0.097 \\
\hspace{2em}4 (\%)& 23.99 & 22.51 & 24.19 & 23.66 & 0.042 & 0.015 \\
\hspace{2em}5 - high (\%)& 53.74 & 43.15 & 55.07 & 52.35 & \bf{0.254} & 0.065\\ \hline

\end{tabular}

\end{table}

\begin{table}[H]
\centering
\caption{Comparison of average baseline variables related to substance abuse, personality, and self-perception before and after matching football players to non-sport controls. Before matching, control values are unweighted. After matching, control values are weighted according to the composition of the matched set (see Table~\ref{tab:matched_set_composition}). Standardized differences greater than 0.2 are bolded.}
\label{tab:balance_match3_substance}
\tiny
\begin{tabular}{lcccccc} \hline
~ & \multicolumn{2}{c}{Before Matching} & \multicolumn{2}{c}{After Matching} & \multicolumn{2}{c}{Standardized Differences} \\
 ~ & Football & Non-sport Controls & Football & Non-sport Controls & Before Matching & After Matching \\ \hline \\
Ever tried cigarette smoking (\%)& 53.55 & 57.69 & 49.77 & 49.58 & -0.099 & 0.005 \\
Smoked regularly (\%)& 14.01 & 24.02 & 11.29 & 12.71 &	\bf{-0.321} & -0.046 \\
Ever drunk alcohol & 55.09 & 55.44 & 51.15 & 48.82 & -0.009 & 0.056 \\
How often subject drank alcohol in past year & ~ & ~ & ~ & ~ & ~ & ~ \\
\hspace{2em}Almost every day (\%)& 0.77 & 1.41 & 0.46 & 0.81 & -0.079 & -0.042 \\
\hspace{2em}3 - 5 days/week (\%)& 2.88 & 3.1 & 1.61 & 2.4 & -0.015 & -0.056 \\
\hspace{2em}1 - 2 days/week (\%)& 6.33 & 7.13 & 4.61 & 5.24 & -0.038 & -0.031 \\
\hspace{2em}2 - 3 days/month (\%)& 10.17 & 7.41 & 8.53 & 5.84  & 0.113 & 0.11 \\
\hspace{2em}Once a month or less (\%)& 10.56  & 10.6 & 10.14 & 8.06 & -0.002 & 0.08 \\
\hspace{2em}Once or twice in past year (\%)& 14.4 & 15.48 & 14.98 & 15.79 & -0.036 & -0.027 \\
\hspace{2em}Never (\%)& 9.79 & 9.94 & 10.6 & 10.45 &	-0.006 & 0.006 \\
\hspace{2em} Not applicable (\%)& 45.11 & 44.93 & 49.08 & 51.41 & 0.003 & -0.047 \\
How often subject had 5+ drinks in a row & ~ & ~ & ~ & ~ & ~ & ~ \\
\hspace{2em}Almost every day (\%)& 0.77 & 0.84 & 0.46 & 0.92 & -0.01 & -0.062 \\
\hspace{2em}3 - 5 days/week (\%)& 1.73 & 2.35 & 0.69  & 0.61 & -0.054 & 0.007 \\
\hspace{2em}1 - 2 days/week (\%)& 4.99 & 4.69 & 3.69 & 2.73 & 0.017 & 0.053 \\
\hspace{2em}2 - 3 days/month (\%)& 6.91 & 6.29 & 6.22 & 4.38 & 0.03 & 0.088 \\
\hspace{2em}Once a month or less (\%)& 6.14 & 5.25 & 5.3 & 3.78 & 0.045 & 0.077 \\
\hspace{2em}Once or twice in past year (\%)& 7.29 & 7.79 &	7.14 & 5.09 & -0.022 & 0.093 \\
\hspace{2em}Never (\%)& 16.89 & 17.92 & 16.36 & 20.63 & -0.032 & -0.135 \\
\hspace{2em} Not applicable (\%)& 55.28 & 54.88 & 60.14 & 61.86 & 0.008 & -0.035\\
How much subject agree that: & ~ & ~ & ~ & ~ & ~ & ~\\
\hspace{1em} You never get sad & ~ & ~ & ~ & ~ & ~ & ~ \\
\hspace{2em}Strongly agree (\%)&	4.03 & 5.63 & 3.69 & 6.62 & -0.092 & -0.169 \\
\hspace{2em}Agree (\%)& 15.16 & 13.13 & 14.75 & 15.8 & 0.069 & -0.036 \\
\hspace{2em}Neither agree nor disagree (\%)& 19.77 & 21.67  & 20.28 & 25.55 & -0.056 & -0.156 \\
\hspace{2em}Disagree (\%)&	53.17 & 48.78 & 53.23 & 41.87 & 0.105 & 0.271 \\
\hspace{2em}Strongly disagree (\%)& 7.87  & 10.79 & 8.06 & 10.16 & -0.124 & -0.089 \\
\hspace{1em} You are physically fit & ~ & ~ & ~ & ~ & ~ & ~ \\
\hspace{2em}Strongly agree (\%)&	42.61& 24.95 &	 41.94 & 42.15 & \bf{0.441} & -0.005 \\
\hspace{2em}Agree (\%)& 47.6 & 48.69 & 47.93 & 44.59 & -0.026 & 0.08 \\
\hspace{2em}Neither agree nor disagree (\%)& 8.06 & 17.26 & 8.06 & 10.65 & \bf{-0.358} & -0.101 \\
\hspace{2em}Disagree (\%)&	1.73 & 8.07 & 2.07 & 2.53  & \bf{-0.413} & -0.03 \\
\hspace{2em}Strongly disagree (\%)& 0 & 0.94 & 0 & 0.09 & \bf{-0.216} & -0.02 \\
\hspace{1em} You feel socially accepted & ~ & ~ & ~ & ~ & ~ & ~ \\
\hspace{2em}Strongly agree (\%)&	38.2 & 28.99 & 38.02 & 40.41 & \bf{0.231} & -0.06 \\
\hspace{2em}Agree (\%)& 54.13 & 54.6 & 54.38 & 48.46 & -0.011 & 0.142 \\
\hspace{2em}Neither agree nor disagree (\%)& 6.14 & 11.26 & 6.22 & 9.17  & \bf{-0.231} & -0.133 \\
\hspace{2em}Disagree (\%)&	1.15 (6/521)& 4.78 & 0.92 & 1.73 & \bf{-0.297} & -0.066 \\
\hspace{2em}Strongly disagree (\%)& 0.38 & 0.38 & 0.46 & 0.23 & 0.002 & 0.045 \\
How much subject feels parents care about him & ~ & ~ & ~ & ~ & ~ & ~ \\
\hspace{2em}Not at all (\%)&	0.19 & 0.28 & 0.23 & 0.31 & -0.023 & -0.02 \\
\hspace{2em}Very little (\%)&	0.77 & 0.66 & 0.92 & 0.23 & 0.016 & 0.097 \\
\hspace{2em}Somewhat (\%)& 2.3 & 3.1 & 1.61 & 1.94 & -0.06 & -0.025 \\
\hspace{2em}Quite a bit (\%)& 10.56 & 13.79 & 9.68 & 9.33 & -0.121 & 0.013 \\
\hspace{2em}Very much (\%)& 85.99 & 81.89 & 87.56 & 88.08 & 0.136 & -0.017 \\ \hline
\end{tabular}
\end{table}

\begin{table}[H]
\centering
\caption{Comparison of average baseline variables related to demographics, general health, and life plan before and after matching sport controls (SC) and non-sport controls (NSC). Before matching, non-sport control values are unweighted. After matching, non-sport control values are weighted according to the composition of the matched set (see Table~\ref{tab:matched_set_composition}). Standardized differences greater than 0.2 are bolded.}
\label{tab:balance_match4_health}
\tiny
\begin{tabular}{lcccccc} \hline
~ & \multicolumn{2}{c}{Before Matching} & \multicolumn{2}{c}{After Matching} & \multicolumn{2}{c}{Standardized Differences} \\
 ~ & SC & NSC & SC & NSC & Before Matching & After Matching \\ \hline \\
Age in 2008 (yrs)& 28.94 & 29.13 & 28.94 & 28.96 & -0.108 & -0.016 \\
Weight in 1994-95 (lbs)&149.66&150.21&149.96&149.39&-0.015&0.016 \\
Height in 1994-95 (in)&69.13&68.11&69.06&68.47& \bf{0.253} &0.147 \\
Subject's reported race & ~ & ~ & ~ & ~ & ~ & ~ \\
\hspace{2em}White (\%)&71.64 & 73.08 &71.94 &74.88 &-0.037 & -0.075 \\
\hspace{2em}Black (\%)& 21.97 & 15.29 &21.58 &17.22 &0.193&0.126 \\
\hspace{2em}Native American (\%)&1.8 & 3.1 &1.8 &2.23 & -0.1&-0.034 \\
\hspace{2em}Asian (\%)& 5.74 & 4.32 & 5.4 &3 & 0.073 &0.123 \\
\hspace{2em}Other (\%)& 3.44 & 7.88 & 3.24 &6.41 & \bf{-0.236} &-0.169 \\
Subject's assessment of his general health & ~ & ~ & ~ & ~ & ~ & ~ \\
\hspace{2em}Excellent (\%)&	37.05 &27.39 &37.05 &35.25 &\bf{0.235} & 0.044 \\
\hspace{2em}Very good (\%)& 42.3 &37.99 &42.81 &38.76 &0.101 &0.095 \\
\hspace{2em}Good (\%)&17.21 &28.8 &16.91 &22.15 &\bf{-0.328}&-0.149 \\
\hspace{2em}Fair (\%)&	2.95 &5.25&2.88 &3.61&-0.139&-0.044 \\
\hspace{2em}Poor(\%)&	0.33&0.56 &0.36 &0.22 &-0.042 &0.024 \\
How often past week subject exercised & ~ & ~ & ~ & ~ & ~ & ~ \\
\hspace{2em}Not at all (\%)&	20.33 & 24.86 & 21.04 & 23.23 & -0.126 & -0.061 \\
\hspace{2em}1 - 2 times (\%)& 28.52 & 32.08 & 27.52 & 31.53 & -0.089 & -0.101 \\
\hspace{2em}3 - 4 times (\%)& 21.48& 22.8 & 21.58 & 23.53 & -0.037 & -0.054 \\
\hspace{2em}5+ times (\%)&	29.51 & 20.26 & 29.86 & 21.7 & \bf{0.242} & 0.213 \\
How often subject had & ~ & ~ & ~ & ~ & ~ & ~ \\
\hspace{1em} Headaches & ~ & ~ & ~ & ~ & ~ & ~ \\
\hspace{2em}Never (\%)& 11.64 & 11.44 & 11.51 & 11.81 & 0.007 & -0.011 \\
\hspace{2em}A few times (\%)& 65.9 & 67.92 & 66.73 & 69.71 & -0.049 & -0.073 \\
\hspace{2em}Once a week (\%)& 20.16 & 16.14 & 19.6 & 15.33 & 0.119 & 0.126 \\
\hspace{2em}Almost every day (\%)& 1.8 & 3.66 & 1.98 & 2.67 & -0.138 & -0.051 \\
\hspace{2em}Every day (\%)& 0.33 & 0.84 & 0.18 & 0.48 & -0.084 & -0.049 \\
\hspace{1em} Unexplained physical weakness & ~ & ~ & ~ & ~ & ~ & ~ \\
\hspace{2em}Never (\%)& 62.62 & 59.19 & 63.13 & 65.02 & 0.081 & -0.045 \\
\hspace{2em}A few times (\%)& 31.8 & 34.62 & 31.47 & 31.26 & -0.069 & 0.005 \\
\hspace{2em}Once a week (\%)& 4.43 & 4.78 & 4.32 &	3.37 & -0.02 & 0.052 \\
\hspace{2em}Almost every day (\%)& 0.49 &	1.22 (13/1066)& 0.54 & 0.34 & -0.097 & 0.026 \\
\hspace{2em}Every day (\%)& 0.49 & 0.19 &	0.54 & 0 &	 0.056 & 0.1 \\ 
\hspace{1em} Dizziness & ~ & ~ & ~ & ~ & ~ & ~ \\
\hspace{2em}Never (\%)& 64.92 & 65.38 & 67.09 & 70.53 & -0.011 & -0.083 \\
\hspace{2em}A few times (\%)& 30.98 & 28.61 & 29.68 & 24.42 & 0.059 & 0.132 \\
\hspace{2em}Once a week (\%)& 2.62 &4.22 & 2.16 & 4.02 &  -0.105 & -0.122 \\
\hspace{2em}Almost every day (\%)& 0.98 & 1.03 & 0.72 & 0.97 &	-0.006 & -0.029 \\
\hspace{2em}Every day (\%)& 0.33 & 0.75 & 0.36 & 0.06 & -0.071 & 0.05 \\
\hspace{1em} Muscle or joint aches & ~ & ~ & ~ & ~ & ~ & ~ \\
\hspace{2em}Never (\%)& 15.57 &	22.8 & 16.37 & 20.04 & \bf{-0.216} & -0.11 \\
\hspace{2em}A few times (\%)& 53.44 &	53.28 & 53.78 & 54.9 & 0.004 & -0.026 \\
\hspace{2em}Once a week (\%)& 21.8 & 19.61 & 21.76 & 20.38 & 0.062 & 0.039 \\
\hspace{2em}Almost every day (\%)& 7.87 & 3.38 & 6.83 & 3.6 & \bf{0.212} & 0.153 \\
\hspace{2em}Every day (\%)& 1.15 & 0.94 & 1.26& 1.08 & 0.023 & 0.02 \\
\hspace{1em} Trouble sleeping& ~ & ~ & ~ & ~ & ~ & ~ \\ 
\hspace{2em}Never (\%)& 44.26 & 42.21 & 45.5 & 46.42& 0.047 & -0.021 \\
\hspace{2em}A few times (\%)& 35.74 &	34.05 & 35.61 & 35.81 & 0.041 & -0.005 \\
\hspace{2em}Once a week (\%)& 11.97 & 13.32 & 11.51 & 11.38 & -0.047 & 0.005 \\
\hspace{2em}Almost every day (\%)& 6.72 & 7.22 & 6.29 & 4.66 & -0.023 & 0.074 \\
\hspace{2em}Every day (\%)& 1.15 & 3.19 & 1.08 & 1.74 & -0.174 & -0.056 \\
Seriously considered suicide in past year (\%) & 7.87 & 11.54 & 7.37 & 7.93 & -0.146 & -0.022 \\ 
Desire to go to college & ~ & ~ & ~ & ~ & ~ & ~ \\
\hspace{2em}1 - low (\%) & 2.95 &	6.85 & 2.7 & 3.88 & \bf{-0.222} &-0.067 \\
\hspace{2em}2 (\%) & 2.95 &	4.88 & 3.06 & 3.12 &	-0.119 &-0.004 \\
\hspace{2em}3 - medium (\%) & 9.67 & 13.6 & 10.25 & 10.69 & -0.144 &-0.016 \\
\hspace{2em}4 (\%) &11.64 & 17.92 & 11.15 &17.25 & \bf{-0.21} & \bf{-0.204} \\
\hspace{2em}5 - high (\%) & 72.46 & 56.57 & 72.66 & 64.94 & \bf{0.393} & 0.191 \\
Likelihood of going to college & ~ & ~ & ~ & ~ & ~ & ~ \\
\hspace{2em}1 - low (\%) & 4.1 & 7.69 & 3.96 & 4.9 & -0.184 & -0.048 \\
\hspace{2em} 2 (\%) &4.43 & 7.5 & 4.68 & 4 & -0.156 & 0.034 \\
\hspace{2em}3 - medium (\%)11.31 & 18.95 & 10.43 & 16.41 & \bf{-0.255} & -0.199 \\
\hspace{2em}4 (\%) &25.41 & 22.51 & 25.36 & 23.16  & 0.077 & 0.059 \\
\hspace{2em}5 - high (\%) & 54.26 &43.15 & 55.22&51.41& \bf{0.256} & 0.088\\ \hline
\end{tabular}

\end{table}

\begin{table}[H]
\centering
\caption{Comparison of average baseline variables related to substance abuse, personality, and self-perception before and after matching sport controls (SC) and non-sport controls (NSC). Before matching, non-sport control values are unweighted. After matching, non-sport control values are weighted according to the composition of the matched set (see Table~\ref{tab:matched_set_composition}). Standardized differences greater than 0.2 are bolded.}
\label{tab:balance_match4_substance}
\tiny
\begin{tabular}{lcccccc} \hline
~ & \multicolumn{2}{c}{Before Matching} & \multicolumn{2}{c}{After Matching} & \multicolumn{2}{c}{Standardized Differences} \\
 ~ & SC & NSC & SC & NSC & Before Matching & After Matching \\ \hline \\
Ever tried cigarette smoking (\%) & 53.28 & 57.69 & 51.98 & 50.88 & -0.102 & 0.025 \\
Smoked regularly (\%) & 14.26 & 24.02 &	12.77 & 15.21&	\bf{-0.296} & -0.074 \\
Ever drank alcohol (\%) & 54.26 & 55.44  & 51.8 & 50.19 & -0.027 & 0.037 \\
How often subject drank alcohol in past year & ~ & ~ & ~ & ~ & ~ & ~ \\
\hspace{2em}Almost every day (\%)& 0.82 & 1.41 & 0.54 & 0.72 & -0.067 & -0.021 \\
\hspace{2em}3 - 5 days/week (\%)& 3.61 & 3.1 & 3.42 & 2.71 &	 0.032 & 0.044 \\
\hspace{2em}1 - 2 days/week (\%)& 6.39 & 7.13 & 5.94 & 4.81 & -0.034 & 0.052 \\
\hspace{2em}2 - 3 days/month (\%)& 6.56 & 7.41& 6.29 & 5.8 & -0.039 & 0.022 \\
\hspace{2em}Once a month or less (\%)&11.97 &10.6 &	11.87 &10.28 &	0.049 & 0.057 \\
\hspace{2em}Once or twice in past year (\%)&14.1&15.48 &13.13 &15.81 &-0.045 & -0.087 \\
\hspace{2em}Never (\%)&10.49&9.94&10.25 &9.61 &0.021 & 0.024 \\
\hspace{2em} Not applicable (\%)&46.07&44.93&48.56 &50.25 &0.023 & -0.034 \\
How often subject had 5+ drinks in a row & ~ & ~ & ~ & ~ & ~ & ~ \\
\hspace{2em}Almost every day (\%)&0.49 &0.84 &0.36 &0.63 &-0.052 & -0.04 \\
\hspace{2em}3 - 5 days/week (\%)&2.46 &2.35&2.52&0.91 &0.008 & 0.12 \\
\hspace{2em}1 - 2 days/week (\%)& 5.41&4.69 &5.4 &3.9 &0.037 & 0.078 \\
\hspace{2em}2 - 3 days/month (\%)&3.61 &6.29&3.06&4.5&-0.148 & -0.08 \\
\hspace{2em}Once a month or less (\%)& 6.23 &5.25& 5.94 & 4.72 & 0.048 & 0.059 \\
\hspace{2em}Once or twice in past year (\%)& 9.51 & 7.79 & 8.81 & 5.7 & 0.069 & 0.126 \\
\hspace{2em}Never (\%)& 15.57 & 17.92 & 14.93 & 19.78 & -0.073 & -0.151 \\
\hspace{2em} Not applicable (\%)& 56.72 & 54.88 & 58.99 & 59.86 & 0.037 & -0.018 \\
How much subject agrees that & ~ & ~ & ~ & ~ & ~ & ~ \\
\hspace{1em} You never get sad & ~ & ~ & ~ & ~ & ~ & ~ \\
\hspace{2em}Strongly agree (\%)&	4.92 & 5.63 & 5.04 & 6.29 & -0.037 & -0.065 \\
\hspace{2em}Agree (\%)& 15.25 & 13.13 & 16.37 & 15.9  & 0.069 & 0.015 \\
\hspace{2em}Neither agree nor disagree (\%)& 19.51 & 21.67 & 19.42 & 21.97 & -0.062 & -0.073 \\
\hspace{2em}Disagree (\%) & 51.31 & 48.78 & 50.54 & 45.97 & 0.058 & 0.105 \\
\hspace{2em}Strongly disagree (\%)& 8.85 & 10.79 & 8.63 & 9.86 & -0.076 & -0.048 \\
\hspace{1em} You are physically fit & ~ & ~ & ~ & ~ & ~ & ~ \\
\hspace{2em}Strongly agree (\%)&	42.62 &24.95 &	41.37 &37.17 & \bf{0.428} & 0.102 \\
\hspace{2em}Agree (\%)&45.41 &48.69&46.22 &48.89 &-0.075 & -0.061 \\
\hspace{2em}Neither agree nor disagree (\%)&9.02&17.26 &9.35&11.11 & \bf{-0.295} & -0.063 \\
\hspace{2em}Disagree (\%)&	2.46 &8.07&2.7&2.58&\bf{-0.317} & 0.007 \\
\hspace{2em}Strongly disagree (\%)& 0.33 & 0.94 &0.36&0.25&-0.096 & 0.016 \\
\hspace{1em} You feel socially accepted & ~ & ~ & ~ & ~ & ~ & ~ \\
\hspace{2em}Strongly agree (\%)&	36.72&28.99& 37.23 & 34.92&	0.188 & 0.056 \\
\hspace{2em}Agree (\%)&54.26 & 54.6 & 54.32 & 54.3 & -0.008 & 0 \\
\hspace{2em}Neither agree nor disagree (\%)& 6.72 & 11.26 & 6.29 & 8.11 &	-0.189 & -0.076 \\
\hspace{2em}Disagree (\%)& 1.97 & 4.78 & 2.16 & 2.49 & -0.192 & -0.022 \\
\hspace{2em}Strongly disagree (\%) & 0 & 0.38 & 0 & 0.18 & -0.12 & -0.058 \\
How much subject feels parents care about him & ~ & ~ & ~ & ~ & ~ & ~ \\
\hspace{2em}Not at all (\%)&	0.49 & 0.28 & 0.36 &	0.27 & 0.037& 0.016 \\
\hspace{2em}Very little (\%)&	0.49 & 0.66 & 0.54 &	0.37 & -0.026 & 0.026 \\
\hspace{2em}Somewhat (\%)& 1.8 & 3.1 & 1.62 & 1.69 & -0.1 & -0.006 \\
\hspace{2em}Quite a bit (\%)& 11.48 & 13.79  & 11.15 & 11.48 & -0.081 & -0.012 \\
\hspace{2em}Very much (\%)& 85.41 & 81.89 & 86.15 & 85.85 & 0.111 & 0.009\\ \hline
\end{tabular}
\end{table}

\end{document}